\documentstyle[12pt]{article}

\newcommand{\p}{\bot}

\newcommand{\dd}{\partial}

\newcommand{\de}{\delta}
\newcommand{\De}{\Delta}
\newcommand{\om}{\omega}
\newcommand{\Om}{\Omega}
\newcommand{\e}{\varepsilon}
\newcommand{\f}{\varphi}
\newcommand{\ls}{\left(}

\newcommand{\rs}{\right)}

\newcommand{\g}{\gamma}
\newcommand{\al}{\alpha}
\newcommand{\be}{\beta}
\newcommand{\ta}{\tau}
\newcommand{\n}{\nu}
\newcommand{\m}{\mu}
\newcommand{\s}{\sigma}
\newcommand{\La}{\Lambda}
\newcommand{\la}{\lambda}

\newcommand{\ra}{\rangle}

\newcommand{\et}{\eta}

\newcommand{\te}{\theta}

\newcommand{\si}{{\rm sign}}
\newcommand{\ti}{\tilde}
\newcommand{\str}[1]{\mathrel{\mathop{\longrightarrow}\limits_{#1}}}

\newcounter{form}

\newcommand{\disn}[2]{$$\displaylines{\refstepcounter{form}
            \label{#1} \hfill #2}$$}
\newcommand{\no}{\hfill \phantom{(\theform)}\cr \hfill}
\newcommand{\nom}{\hfill (\theform) \cr}

\newcounter{punkt}
\makeatletter
\makeatletter

\newcount\dobav
\newcount\otlog
\newcount\vrem
\def\@citex[#1]#2{\if@filesw\immediate\write\@auxout{\string\citation{#2}}\fi
  \let\@citea\@empty
  \dobav=-1
  \otlog=-1
  \@cite{\@for\@citeb:=#2\do
    {\def\@tempa##1##2\@nil{\edef\@citeb{\if##1\space##2\else##1##2\fi}}%
     \expandafter\@tempa\@citeb\@nil
     \@ifundefined{b@\@citeb}{\@warning%
       {Citation `\@citeb' on page \thepage \space undefined}%
       \vrem=-1}{\vrem=\csname b@\@citeb\endcsname}
\advance\vrem by -1
\ifnum \vrem=\dobav
 \otlog=\vrem
 \advance\otlog by 1
\else
 \ifnum \vrem=\otlog
  \advance\otlog by 1
 \else
  \ifnum \otlog>0
   \advance\dobav by 1
   \ifnum \otlog=\dobav
    \hbox{,\penalty\@m\ \the\otlog}%
   \else
    \hbox{--\the\otlog}%
   \fi
   \otlog=-1
  \fi
  \dobav=\vrem
  \advance\dobav by 1
  \@citea\def\@citea{,\penalty\@m\ }%
  \ifnum \dobav=-1
   {\reset@font\bf ?}%
  \else
   \hbox{\the\dobav}%
  \fi
 \fi
\fi
}%
\ifnum \otlog>0
 \advance\dobav by 1
 \ifnum \otlog=\dobav
  \hbox{,\penalty\@m\ \the\otlog}%
 \else
  \hbox{--\the\otlog}%
 \fi
\fi
}{#1}}

\renewcommand{\section}{\@startsection{section}{1}{0pt}%
            {3.5ex plus 1ex minus .2ex}{2.3ex plus .2ex}{\bf}}
\long\def\@makecaption#1#2{%
   \vskip 10\p@
   \setbox\@tempboxa\hbox{#1. #2}%
   \ifdim \wd\@tempboxa >\hsize
       #1. #2\par
     \else
       \hbox to\hsize{\hfil\box\@tempboxa\hfil}%
   \fi}
\makeatother
\newcommand{\sect}[2]{\protect\refstepcounter{punkt}\protect\label{#1}
            \section*{$\protect\vphantom{a}$\hfill
            \arabic{punkt}.\hskip 2mm #2 \hfill $\protect\vphantom{a}$}}
\newcommand{\st}{\hfill $\protect\vphantom{a}$\protect\\
            $\protect\vphantom{a}$\hfill}

\begin{document}
\large

\title{On the construction of corrected Light-Front Hamiltonian for $QED_2$}

\author{S.~A.~Paston\thanks{E-mail: Sergey.Paston@pobox.spbu.ru}, 
E.~V.~Prokhvatilov\thanks{E-mail: Evgeni.Prokhvat@pobox.spbu.ru}, 
V.~A.~Franke\thanks{E-mail: franke@snoopy.phys.spbu.ru}\\
St.-Peterspurg State University, Russia}

\date{November 23, 2000}

\maketitle

\begin{abstract}
The   counterterms,  which  must  be  included  into  Light-Front
Hamiltonian  of  $QED_2$ to get the equivalence with conventional
Lorentz-covariant  formulation,  are found. This is done to
all  orders  of  perturbation  theory  in fermion mass, using the
bosonization  at intermediate steps and comparing Light-Front and
Lorentz-covariant  perturbation theories for bosonized model. The
obtained  Light-Front  Hamiltonian contains all terms, present in
the   $QED_2$  theory, canonically  (naively)  quantized  on  the
Light-Front   (in   the   Light-Front   gauge)   and  an  unusual
counterterm.   This   counterterm   is   proportional  to  linear
combination  of  fermion  zero modes (which are multiplied by some operator
factors neutralizing their charge and fermionic number). The coefficients before
these zero mode operators are UV finite and depend on condensate parameter in the
$\theta$-vacuum.  These
coefficients  are  proportional  to fermion mass, when this mass goes to
zero.
\end{abstract}

\newpage
\sect{vved}{Introduction}

Hamiltonian approach  to Quantum Field Theory in Light-Front (LF) coordinates
\cite{1} $x^{\pm}=(x^0 \pm x^3)/\sqrt{2},\; x^{\p}= (x^1,x^2) $, with $x^+$
playing the role of time, is one of nonperturbative approaches which can be used
in attempts to solve strong coupling problems \cite{2,3,4}. One quantizes field
theory
on the hyperplane $x^+=0$ and considers the generator $P_+$ of the shift along
the $x^+$ axis  as a Hamiltonian $H$. The generator of the shift along the
axis $x^-$, i.~e. the momentum operator $P_-$,
does not displace the surface $x^+=0$,
where the quantization is performed, and is kinematic (according to Dirac
terminology) in contrast to dynamical generator $P_+$. Operator $P_-$ does
not depend on the interaction and is quadratic in field variables. Due to 
spectral
condition it is nonnegative and has zero eigenvalue on the physical vacuum.
Positive and negative $p_-$ parts of Fourier modes of fields play a role of
creation and annihilation operators on this vacuum and can be used to form 
LF Fock space.
Therefore the physical vacuum can be trivially described
in terms of this "mathematical" Fock space vacuum.
The spectrum of bound states can be found by solving the Schroedinger equation
 \disn{1}{
P_+|\Psi\rangle =p_+|\Psi\rangle
\nom}
in the subspaces with fixed $p_-,p_{\p}$. The mass $m$ can be found as
$m^2= 2p_+p_- -p_{\p}^2$. The procedure of solving this bound state problem
can be nonperturbative. This was demonstrated for (1+1)-dimensional field theory
models by so called DLCQ method \cite{5,5',6} (see  \cite{3}).

LF Hamiltonian formalism faces with specific divergences at $p_-=0$
\cite{2,3,4,7}, and  needs
a regularization. Simplest translationally invariant regularization is the
cutoff  $p_-\ge\e >0$. This cutoff breaks Lorentz and gauge symmetries.
Another regularization, retaining gauge invariance,   is  the cutoff
 $|x^-|\le L$  with periodic boundary conditions for fields.
In this case the momentum $ p_-$
is discrete, $p_-=p_n=(\pi n/L),\; n=0,1,2,...$,
and zero modes, $p_-=0$, are well
separated. Canonical
formalism allows in principle to express these zero modes in terms
of other, nonzero modes via solving constraints (usually it is a very
complicated nonlinear problem) \cite{8,9,10,11,12}.

Examples of nonperturbative calculations in (1+1)-dimensional models,
using canonical LF formulation, show that the description of vacuum
effects can be nonequivalent to that in the conventional formulation in Lorentz
coordinates \cite{13,14,15}.
Furthermore lowest order perturbative calculations  show a difference
between corresponding LF and Lorentz-covariant Feynman diagrams \cite{16,17}.
This can invalidate the conventional renormalization procedure for
LF perturbation theory.
These violations of the equivalence between Lorentz-covariant and canonical 
LF formulations can be
caused by the breakdown of Lorentz  and other symmetries due to LF
regularizations described above. Nevertheless for nongauge theories, like
Yukawa model, it is possible to restore the equivalence, at least perturbatively,
by adding few simple counterterms to canonical LF Hamiltonian
\cite{17,18,19,20}.
For gauge theories we have used the general
method of the paper \cite{20}, to compare the Feynman diagrams of LF and 
conventional (covariant)
perturbation theories (to all orders) and to construct via this method
counterterms for canonical LF Hamiltonian. It appears that the number of
such counterterms under simple $|p_-|\ge\e$ regularization
is infinite \cite{20}. We are able to
overcome this difficulty only via a complication of the regularization scheme
(using Pauli-Villars type "ghost" fields) \cite{21}.
The theory on the LF with the resulting $QCD$-Hamiltonian of \cite{21} is
perturbatively equivalent in the limit of removing the regularization
to the Lorentz-covariant $QCD$. But this LF
Hamiltonian is rather complicated and contains several unknown
parameters. Furthermore it does not guarantee that the
description of nonperturbative vacuum effects will be correct.

One can explicitly analyse these vacuum effects only in
two-dimensional gauge models. Inspite of the simplicity of 2-dimensional formulation
  one may hope to get some indications 
how to describe the nonperturbative vacuum effects also in four dimensions.

In (1+1)-dimensional space-time it is
possible to go beyond usual perturbation theory by transforming
gauge field model, like $QED_2$, to scalar field model ("massive
Sine-Gordon" type model for the  $QED_2$) \cite{22',22,23,24}. This transformation
includes the
"bosonization" procedure, i.~e. the transition from fermionic to bosonic
variables \cite{15,25,26,27,28}.
After such transformation of the Hamiltonian the
fermionic mass term
plays the role of the interaction term in corresponding scalar field theory.
Therefore  the perturbation theory in terms of this scalar field (with
fermion mass becoming the coupling constant) plays the role of conventional
chiral perturbation theory.   Furthermore  nontrivial
description of quantum vacuum in $QED_2$,
related with instantons and the "$\theta$"
vacua \cite{23,24,28',adam}, is taken into account in  this scalar field theory
simply by the appearing of the "$\theta$"--parameter in the interaction term.

In the present paper we use this formulation to construct  correct LF
$QED_2$ Hamiltonian, applying perturbation theory ( to all orders in coupling, i.e. in 
fermion mass)
 in corresponding scalar field  theory. After obtaining corrected boson form 
of LF Hamiltonian we make the transformation to original
canonical LF form (with fermionic field variables on the LF) plus some new terms
("counterterms").

Due to nonpolynomial character of boson field interaction,  in perturbation theory
one faces ultraviolet (UV) divergencies for some infinite
sums of
diagrams of a given order \cite{30}.
That is why it is difficult to apply straightforwardly the method of the paper \cite{20}
of the comparison of LF and covariant perturbation theories. In the paper
\cite{30} we used Pauli-Villars regularization for scalar
field.  This allowed to apply the general
method of \cite{20} and to find the
counterterms that correct the boson LF Hamiltonian \cite{30}
(analogous consideration for sine-Gordon model not containing
UV-divergences was carried out in the paper \cite{29}). These
counterterms are proportional to chiral condensate parameter, which
depends on the UV cutoff and becomes infinite after removing  the
regularization (while perturbatively one can see how this infinity cancels the
divergencies appearing in LF perturbation theory).
The LF Hamiltonian corrected in this way depends explicitly on the
UV cutoff parameter. Furthermore
we can not return simply to original canonical LF variables
(the transformation from boson
to fermion variables on the LF becomes complicated due to the presence of boson
Pauli-Villars ghost fields in the chosen regularization scheme).

In the present paper we avoid these difficulties
using, instead of boson Pauli-Villars UV regularization, simple
UV cutoff at intermediate steps.
This becomes possible, because we are able to prove UV finiteness 
of Lorentz-covariant Green functions of the considered scalar
field theory (without vacuum loops)
to all orders of perturbation theory in fermion mass parameter.
We carry out a detailed comparison of LF and Lorentz-covariant
perturbation theories to all orders. As a result,
we find counterterms to canonical LF Hamiltonian restoring its
equivalence with Lorentz-covariant formulation (in terms of boson field).
The form of the obtained LF Hamiltonian allows
simple transformation to original canonical fermion variables on the LF and
the direct comparison with the original canonical LF Hamiltonian.

The resulting form of LF Hamiltonian includes a term, which appears
under naive canonical LF quantization of $QED_2$ in LF gauge
and a new counterterm proportional to linear
combination  of  fermion  zero modes (multiplied by some operator
factors neutralizing their charge and fermionic number) with the
UV finite coefficients, depending on condensate parameter. These
coefficients  are  proportional  to fermion mass, when this mass approaches to
zero.

One can use this LF Hamiltonian for nonperturbative calculations
applying DLCQ
method and fitting condensate parameters to known spectrum.

However it is more interesting to extract from this study some features
that could be true also in the 4-dimensional $QCD$.
This was attempted recently in  \cite{mac1,mac2}.

In Sect.~2 we review Hamiltonian formulation of $QED_2$ and the canonical
form of the bozonization transformation. In Sect.~3  we prove the
UV finiteness of the bosonized Lorentz-covariant perturbation
theory to all orders in fermion
mass. In Sect.~4-6 we compare LF and Lorentz-covariant perturbation
theories in boson form and construct counterterms generating
the difference between them. At the beginning of the Sect.~4
a more detailed explication of the content of the Sect.~4-6 is 
given. In Sect.~7 we rewrite the corrected  boson LF
Hamiltonian in terms of canonical fermion variables on the LF.

\sect{boz}{Hamiltonian formulation of $QED_2$ in Lorentz frame \st
and the bosonization transformation}

Let us start from usual Lagrangian density for $QED_2$ in Lorentz coordinates
$x^{\mu}=(x^0,x^1)$:
 \disn{2.2}{
L=-\frac{1}{4}F_{\m\n}F^{\m\n}+\bar\Psi(i\g^mD_\m-M)\Psi,
\nom}
where $F_{\mu\nu}= \dd_{\mu} A_{\nu}- \dd_{\nu }A_{\mu}$,
$D_{\mu}=\dd_{\mu} - ieA_{\mu} $,  $A_{\mu}(x)$  are gauge fields,
$\Psi = {\psi_- \choose \psi_+}$, $\bar\Psi =\Psi^+\g^0$
are fermion fields with mass $M$, $e$ is the
coupling constant, $\gamma^{\mu}$ are chosen as follows:
 \disn{2.2.1}{
\g^0=\ls
\begin{array}{cc}
0 & -i\\
i & 0
\end{array}
\rs, \quad \g^1=\ls
\begin{array}{cc}
0 & i\\
i & 0
\end{array}
\rs, \quad \g^5=i\g^0\g^1=\ls
\begin{array}{cc}
i & 0\\
0 & -i
\end{array}
\rs.
\nom}

From this Lagrangian one gets the following Hamiltonian $H$:
 \disn{2.3}{
H=\int dx^1\ls\frac{1}{2}\Pi_1^2+\sum_{r=\pm}r\ls \psi^+_riD_1\psi_r\rs+
iM\ls\psi_+^+\psi_--\psi_-^+\psi_+\rs\rs,
\nom}
where $\Pi_1=\frac{\dd L}{\dd(\dd_0A_1)}= F_{01}$
is the momentum conjugate to $A_1$.
Besides one gets the constraint equation
 \disn{2.4}{
\dd_1\Pi_1+e\ls\psi^+_+\psi_++\psi_-^+\psi_-\rs=0.
\nom}

It is convenient to consider the theory on the
interval $-L \le x^1 \le L $ with
periodic boundary conditions for fields and to fix the appropriate gauge
\cite{9,10,11,12}:
 \disn{2.5}{
\dd_1A_1=0.
\nom}
Then all fields can be represented by corresponding Fourier series,
and zero, $p_1=0$, mode can be separated from nonzero modes.
We denote the zero and nonzero mode parts of a function $f(x^1)$
as follows:
 \disn{2.6}{
f(x^1)\equiv f_{(0)}+[f(x^1)], \quad
f_{(0)}\equiv \int\frac{dx^1}{2L}f(x^1).
\nom}
Owing to the constraint (\ref{2.4}) we get
 \disn{2.7}{
[\Pi_1]=-e\dd_1^{-1}[\psi^+_+\psi_++\psi^+_-\psi_-],
\nom}
 \disn{2.8}{
Q\equiv \int\limits_{-L}^Ldx^1\ls\psi^+_+\psi_++\psi^+_-\psi_-\rs=0.
\nom}
We use eq-n (\ref{2.7}) to exclude $[\Pi_1]$ from the 
Hamiltonian (\ref{2.3}).
The zero modes $A_1\equiv A_{1(0)}$, $\Pi_{1(0)}$ remain
independent and commute at fixed time as follows:
 \disn{2.8.1}{
[A_{1(0)},\Pi_{1(0)}]=\frac{i}{2L}.
\nom}
Here and in the following we assume that
$(\dd_1^{-1}[f(x^1)])_{(0)}\equiv 0$.
We impose the condition (\ref{2.8}) on physical states after quantization:
 \disn{2.9}{
Q|phys\rangle=0.
\nom}

Quantum definition of the charge $Q$ and current operators
$I_{\pm}(x)=$\linebreak $\psi_{\pm}^+(x)\psi_{\pm}(x)$
includes gauge invariant UV regularization \cite{15,26}
(the "point splitting" with the parameter $\delta$) of the following form:
 \disn{2.10}{
I_r(x;\de)=\psi^+_r(x^1-\frac{ri\de}{2})
\psi_r(x^1+\frac{ri\de}{2})\exp(reA_1\de),\quad r=\pm.
\nom}
Here and in the following we set $x^0=0$ and
omit this variable for brevity.
 \disn{2.11}{
Q(\de)=\int\limits_{-L}^Ldx^1\sum_{r=\pm}I_r(x;\de).
\nom}
This regularization makes the Hamiltonian well defined \cite{26}.
Indeed, the term  of the Hamiltonian (\ref{2.3}),
 \disn{2.12}{
H_{\bar\Psi D\Psi}(\de)=\!\int\limits_{-L}^Ldx^1\sum_{r=\pm}
r\ls\psi^+_r(x^1-\frac{ri\de}{2}) iD_1\psi_r(x^1+\frac{ri\de}{2})
\exp(reA_1\de)\rs=\no
=\frac{d}{d\de}Q(\de),
\nom}
can be minimized by filling negative "energy" levels (according to Dirac
procedure). Following the work \cite{26} one can find
the minimizing state, starting from the state vector $|0_D\rangle$
such that
 \disn{2.14}{
\psi_r(x)|0_D\rangle=0,
\nom}
using Fourier decomposition
 \disn{2.13}{
\psi_r(x)=\frac{1}{\sqrt{2L}}\sum_{n=-\infty}^\infty\psi_{n,r}e^{-ip_nx^1},
\quad p_n=\frac{\pi}{L}n
\nom}
and introducing a set of states $|q\ra=|q_+,q_-\ra$ with "energy" levels
filled up to $q_{\pm}$:
 \disn{2.15}{
|q\ra=\prod_{n=-\infty}^{q_+}\prod_{m=q_-}^\infty
\psi_{n,+}^+\psi_{m,-}^+|0_D\ra,
\nom}
where $q_+$, $q_-$ are integers and
 \disn{2.15.1}{
[\psi_{n,r}^+,\psi_{m,r'}]_+=\de_{nm}\de_{rr'}.
\nom}
One gets \cite{15,26}
 \disn{2.16}{
\int\limits_{-L}^{L}dx^1I_r(x,\de)|q\ra=
\ls\frac{L}{\pi\de}+\ls rq_r+\frac{1}{2}+\frac{rLeA_1}{\pi}\rs\rs
|q\ra+O(\de),
\nom}
 \disn{2.17}{
H_{\bar\Psi D\Psi}(\de)|q\ra=\no=
\ls -\frac{2L}{\pi\de^2}-\frac{\pi}{12 L}+\frac{\pi}{2L}
\sum_{r=\pm}\ls rq_r+\frac{1}{2}+\frac{rLeA_1}{\pi}\rs^2\rs|q\ra+O(\de).
\nom}
These expressions show that
one can define the renormalized (i.~e. UV finite in the
$\delta\to 0$ limit) quantities as follows:
 \disn{2.18}{
I_r(x)=\lim_{\de\to 0}\ls I_r(x;\de)-\frac{1}{2\pi\de}\rs,
\nom}
 \disn{2.19}{
Q=\lim_{\de\to 0}\ls Q(\de)-\frac{2L}{\pi\de}\rs=
\int\limits_{-L}^Ldx^1\sum_{r=\pm}I_r(x),
\nom}
 \disn{2.20}{
H_{\bar\Psi D\Psi}=\lim_{\de\to 0}\ls H_{\bar\Psi D\Psi}(\de)+
\frac{2L}{\pi\de^2}+\frac{\pi}{12 L}\rs.
\nom}
Moreover one can define the operators $Q_{\pm}$:
 \disn{2.21}{
Q_r\equiv \ls\int\limits_{-L}^Ldx^1I_r(x)\rs-\frac{1}{2}-\frac{rLeA_1}{\pi},
\qquad Q_++Q_-=Q-1,
\nom}
 \disn{2.22}{
Q_r|q\ra=rq_r|q\ra.
\nom}
According to the equation (\ref{2.16}) (or (\ref{2.22})) and the paper \cite{26}
these operators are integer 
valued and commute with gauge fields and current operators:
 \disn{2.23}{
[Q_r,\Pi_{1(0)}]=[Q_r,A_1]=[Q_r,I_{r'}(x)]=0
\nom}
(let us remark that applying current operators to states $|q\ra$,
we can get the whole Hilbert space \cite{uhl}).
This definition leads to the commutation relation between $\Pi_1$ and
$Q_5=\int_{-L}^{L}dx^1(I_+ -I_-)$:
 \disn{2.24}{
[Q_5,\Pi_{1(0)}]=\frac{ie}{\pi}
\nom}
because of eq-n (\ref{2.8.1}).
Renormalized currents (\ref{2.18}) satisfy commutation relations 
with anomalous
Schwinger term \cite{13,14,15,26}:
 \disn{2.25}{
[I_r(x),I_{r'}(x')]_{x^0=0}=\frac{ir}{2\pi}\de_{rr'}\dd_1\de(x^1-x'^1).
\nom}
These relations account for the axial vector anomaly in $QED_2$.
On the other hand
they are a base for the bosonization, because their form resembles boson
field canonical commutation relations. Indeed one can introduce the following
canonically conjugate boson field variables:
 \disn{2.26}{
\Phi(x)=-\frac{\sqrt{\pi}}{e}\ls\Pi_{1(0)}-e\dd_1^{-1}[I_++I_-]_x\rs=
-\frac{\sqrt{\pi}}{e}\Pi_1(x),
\nom}
 \disn{2.27}{
\Pi(x)=\sqrt{\pi}\ls I_+(x)-I_-(x)\rs,
\nom}
 \disn{2.28}{
[\Phi(x),\Phi(x')]_{x^0=0}=[\Pi(x),\Pi(x')]_{x^0=0}=0,\no
[\Phi(x),\Pi(x')]_{x^0=0}=i\de(x^1-x'^1).
\nom}
Let us construct inverse transformation formula, that expresses initial
fermion variables in terms of boson ones. With this aim we introduce
\cite{13,14,15,26} operators $\omega_{\pm}$,
canonically conjugated to the $Q_{\pm}$ (and
commuting with other canonical variables):
 \disn{2.29}{
[\om_r,Q_{r'}]=i\de_{rr'}.
\nom}
According to the equality
 \disn{2.30}{
e^{-i\om_r}Q_re^{i\om_r}=Q_r+1,
\nom}
these operators can be represented by their action on the $|q\ra$ states:
 \disn{2.31}{
e^{i\om_+}|q_+,q_-\ra=|q_++1,q_-\ra,\qquad
e^{i\om_-}|q_+,q_-\ra=|q_+,q_--1\ra.
\nom}
The expression for fermion fields can be written in the same way as it is
done in string theory \cite{15,28}:
 \disn{2.32}{
\psi_r(x)=N_0e^{-i\om_r}e^{ri\frac{\pi}{2}Q-ri\pi\frac{x^1}{L}Q_r}
:e^{-i\sqrt{\pi}[\dd_1^{-1}[\Pi]+r\Phi]_x}:,
\nom}
where $N_0$ is a normalization factor depending on the choice of normal
ordering in (\ref{2.32}).
To fix this normal ordering we take Fourier decompositions
of boson fields in the interaction picture form (i.~e. in the
free field like form with the mass $m=e/\sqrt{\pi}$ and
operators $a_n^+, a_n$ playing the role of creation and annihilation
operators):
 \disn{2.33}{
[\Phi(x)]=\frac{1}{\sqrt{4L}}\sum_{n\neq 0}
\frac{a_ne^{-ip_nx^1}+h.c.}{\sqrt{E_n}},
\nom}
 \disn{2.34}{
[\Pi(x)]=\frac{1}{i\sqrt{4L}}\sum_{n\neq 0}
\sqrt{E_n}\ls a_ne^{-ip_nx^1}-h.c.\rs,
\nom}
 \disn{2.34.1}{
[a_m,a_n^+]=\de_{mn},\qquad
E_n\equiv \sqrt{m^2+p_n^2}.
\nom}
Then we take normal ordering with respect to operators $a_n, a_n^+$, and the
factor  $N_0$ becomes equal to
 \disn{2.35}{
N_0=\frac{1}{\sqrt{2L}}\exp\ls -\frac{\pi}{4L}
\sum_{n>0}\frac{(E_n-p_n)^2}{p_n^2E_n}\rs.
\nom}
One can check \cite{15} (taking into account UV regularization
(\ref{2.10}) of the  product
of fermion fields) that operators  (\ref{2.32})  satisfy correct canonical
anticommutation relations and reproduce the definitions
(\ref{2.26}),(\ref{2.27}).

Having the explicit formulas for the bosonization one can rewrite the
Hamiltonian (\ref{2.3})
(where the substitution of (\ref{2.7}) for the $[\Pi_1]$ is done) in
terms of boson variables. One gets on the physical subspace of states which
satisfy constraint (\ref{2.9}), the following expression  \cite{13,14,15}:
 \disn{2.36}{
H=\int\limits_{-L}^Ldx^1:\ls \frac{1}{2}\Pi^2+\frac{1}{2}(\dd_1\Phi)^2+
\frac{m^2}{2}\Phi^2-MN_1\cos(\om+\sqrt{4\pi}\Phi)\rs:+\no
+const,
\nom}
where  $\omega\equiv \omega_+ -\omega_- -\sqrt{4\pi}\Phi_{(0)} - \pi/2$,
$N_1=\frac{1}{L}\exp\ls\frac{\pi}{L}\sum_{n>0}(\frac{1}{p_n}-\frac{1}{E_n})\rs$.
This Hamiltonian coincides with the Hamiltonian of scalar field theory, where
the field $\Phi(x)$ is periodic on the interval
$-L\le x^1 \le L$  and the operator $\omega$, commuting with all boson field
variables, can be replaced by it's eigenvalue, playing the role of
the "$\theta$" -- parameter. The relation
between $\omega$ and the $\theta$-parameter
can be explained by noticing that the remaining (after fixing the gauge
$\dd_1A_1=0$) discrete gauge group
 \disn{2.38}{
\Psi_n\str{\Om_l}\Psi_{n+l},\qquad A_1\str{\Om_l}A_1+\frac{\pi l}{eL}
\nom}
is realized by the operators
$\Omega_l=\exp(il\omega)$, and that these gauge operators are responsible
for the change of topological charge and connected
with the definition of the $\theta$-vacuum \cite{24,adam}.
The operators $a_n$,
determined by eq-ns (\ref{2.33}),(\ref{2.34}),
annihilate free boson field vacuum $|0\ra$:
 \disn{2.37}{
a_n|0\ra=0.
\nom}
This vacuum state can be described also in terms of states (\ref{2.15})
\cite{15,26}.

In the limit $mL\to\infty$ one gets \cite{15}
 \disn{2.39}{
\lim_{mL\to\infty}N_1=\frac{me^C}{2\pi},
\nom}
where $C=0.577216\dots$ is the Euler constant, and the Hamiltonian
(\ref{2.36}) takes the
form corresponding to Lorentz-invariant theory of scalar field $\Phi$:
 \disn{2.41}{
H=\int dx^1:\ls\frac{1}{2}\Pi^2+\frac{1}{2}(\dd_1\Phi)^2+
\frac{m^2}{2}\Phi^2-\frac{Mme^C}{2\pi}\cos(\te+\sqrt{4\pi}\Phi)\rs:.
\nom}
This scalar field theory is considered in the sections~3-6.

\sect{uvfin}{The proof of the UV finiteness of the
Lorentz-covariant \st boson theory
to all orders in fermion mass}

Let us start with Lagrangian corresponding to the Hamiltonian (\ref{2.41}):
 \disn{3.0}{
L=L_0+L_I,
\nom}
 \disn{3.1}{
L_0=\frac{1}{8\pi}\ls\dd_\m\f\dd^\m\f-m^2\f^2\rs,
\nom}
 \disn{3.2}{
L_I=\frac{\g}{2}e^{i\te}:e^{i\f}:+\frac{\g}{2}e^{-i\te}:e^{-i\f}:,\quad
\g=\frac{Mme^C}{2\pi},
\nom}
where for convenience we use the notation $\f=\sqrt{4\pi}\Phi$;
the normal ordering symbol means that in the perturbation theory in $\g$
one excludes all diagrams with lines starting and ending at the 
same vertex
(it also corresponds to usual normal ordering of the operators
in a interaction Hamiltonian).
Let us consider the structure of Feynman
diagrams in Lorentz coordinates for this theory. There are two
types of vertices with $j$  lines, the 1st type giving the factor
$i^{j+1}e^{i\theta}\gamma/2$ and the 2nd giving the factor
$i^{-j+1}e^{-i\theta}\gamma/2$.
The vertex without lines, $j=0$, is considered as a connected diagram.
In the following we subdivide the vertex factors into two parts,
$ie^{{\pm} i \theta}\gamma/2$, related with vertex point, and $({\pm}i)^j$,
related with $j$ corresponding outgoing lines. The propagators
$\Delta (x) = \langle 0|T(\f (x)\f(0))|0\ra$,
where free field $\f(x)$ corresponds
to the decomposition (\ref{3.0}) of the Lagrangian,
can be written as follows:
 \disn{3.4}{
\De(x)=\int d^2k\;e^{ikx}\De(k),\qquad
\De(k)=\frac{i}{\pi}\frac{1}{(k^2-m^2+i0)},
\nom}
where $d^2k=dk_0dk_1$, $kx=k_0x^0+k_1x^1$.

Any pair of vertices in diagrams can be connected by any number of propagators.
As shown in the Appendix~1
(see also \cite{adam}) the perturbation theory can be
reformulated in terms of sums over all such ways to
connect a pair of vertices.
We call these sums "superpropagators"
(clearly this term has nothing to do with supersymmetry).
For vertices of different type the
superpropagator is equal to $e^{\Delta (x)}$, and for vertices of the same
type it is $e^{-\Delta (x)}$. Due to the presence of the
unity in the expansions of these
exponents all "superdiagrams", constructed with superpropagators, are disconnected
in conventional sense. In the following we hold to the conventional definition
of the connectedness.

Consider the sum $S'_n$
of all diagrams of order $n$ in $\g$ with fixed attachment of
external lines and  fixed types of
vertices. This sum can be represented with one superdiagram, where each
pair of vertices is connected by the superpropagator of corresponding type. Let
$S_n$ be the sum of only connected diagrams, contributing to the $S'_n$.
We can write
 \disn{3.6.1}{
S_n=S'_n-S''_n,
\nom}
where the $S''_n$ is corresponding sum of disconnected diagrams.
The $S''_n$  can be considered as a sum of terms which themselves are
sums of all disconnected diagrams with fixed subdivision of the initial
superdiagram into connected parts. Every such term is a product of some
$S_{n_1}$ with $n_1<n$. Now one can write the decomposition (\ref{3.6.1})
for every
$S_{n_1}$ and repeat the procedure up to the $S_1$, which is equal to the
$S'_1$. Thus one can get the representation of the $S_n$ in terms of a
sum of products of the $S'_j$ with $j\le n$:
 \disn{3.6.2}{
S_n=\sum\prod_kS'_{j_k},
\nom}
 \disn{3.6.3}{
\sum_kj_k=n.
\nom}
Let us use the $x$-representation.
We estimate the index of UV divergency of the $S_n$ via simple
dimensional counting in the integrands, considering
all vertex coordinates approaching each other (this estimation doesn't take
into account possible UV divergences of subdiagrams  \cite{wein}).
Having in the integrand of the $S_n$ a pole of order $r'_n$,
one estimates the index as
 \disn{3.11}{
2(n-1)-r'_n,
\nom}
where $2(n-1)$ volume elements in the integrals are taken into account.
To estimate the $r'_n$ one can use the decomposition (\ref{3.6.2}),
applying it to the integrands of diagrams before integration.

Let us find the power $r_j$ of the pole in the integrand of the $S'_j$.
With this aim we represent the mentioned integrands in terms of
superpropagators.
Let a superdiagram $S'_j$ has $l_1$ vertices
of the 1st type and  $l_2$ vertices of the 2nd type, $l_1+l_2=j$. Using the
property of McDonald function $K_0(z)$ for $z\to 0$ we write
 \disn{3.7}{
\De(x)\sim K_0(m\sqrt{x^2})\sim \ln\frac{1}{x^2}.
\nom}
Hence, corresponding superpropagators behave as follows:
 \disn{3.8}{
e^{\pm\De(x)}\sim(x^2)^{\mp 1}.
\nom}
At coinciding coordinates of vertices one can sum up the powers  of $x^2$ for
all superpropagators. The number of superpropagators joining vertices of the same
type is $l_1(l_1-1)/2+l_2(l_2-1)/2$, and the number of those joining vertices
of different type is $l_1l_2$. Therefore one can write
 \disn{3.10}{
r_j=2\ls l_1l_2-\frac{l_1(l_1-1)}{2}-\frac{l_2(l_2-1)}{2}\rs=
j-(l_1-l_2)^2\le j.
\nom}
Now the $r'_n$ can be written as follows
 \disn{3.10.1}{
r'_n=\max\sum_kr_{j_k},
\nom}
where the maximum is taken over the terms in the eq-n (\ref{3.6.2}).
According to eq-ns (\ref{3.6.3}),(\ref{3.10}) one gets
 \disn{3.10.2}{
r'_n\le \max\sum_kj_k=n.
\nom}
Hence for the index (\ref{3.11}) we have
 \disn{3.10.3}{
2(n-1)-r'_n\ge 2(n-1)-n=n-2.
\nom}
The necessary condition for the UV convergence of the $S_n$ is the
positivity of this index, i.~e. the validity of the inequality $n>2$.
At $n=1$ there is no UV
divergency due to the normal ordering prescription in the Lagrangian.
At $n=2$ the eq-n (\ref{3.10}) reads
 \disn{3.10.5}{
r'_2=r_2=2-(l_1-l_2)^2,\qquad l_1+l_2=2,
\nom}
hence, for corresponding index  one has according to eq-n (\ref{3.11})
 \disn{3.10.6}{
2(2-1)-r'_2=(l_1-l_2)^2.
\nom}
The condition of positivity of this index  is $l_1=2,l_2=0$ or $l_1=0,l_2=2$.
This is fulfilled  if both vertices are of the same type. For $l_1=l_2=1$, i.~e.
for the vertices of different type, one may have logarithmic UV divergency for
the $S_2$. However the complete contribution to the connected Green
function in this order is UV finite. Indeed, the corresponding
Green function for the vertices of different type is (see Appendix~1)
 \disn{3.16}{
G_N^{(2)}(y_1,\dots,y_N)=\no=-\ls\frac{\g}{2}\rs^2
\int d^2x_1d^2x_2\prod_{i=1}^N\ls\sum_{k_i=1,2}(-i)(-1)^{k_i}
\De(y_i-x_{k_i})\rs e^{\De(x_1-x_2)},
\nom}
where the sum is over different ways of the attachment of external lines to the
vertices. As  $(x_1-x_2)\to 0$ in the integrand, one has
 \disn{3.17}{
\sum_{k_i=1,2}(-1)^{k_i}\De(y_i-x_{k_i})\sim O(x_1-x_2).
\nom}
Therefore divergent terms cancel in the integral (\ref{3.16}),
and corresponding Green function is UV finite for  $N>0$.

Let us consider  the possibility that the $S_n$ is UV divergent due to
bad UV index of some subdiagrams \cite{wein}.
Such subdiagrams can be only
of 2nd order in $\gamma$ with vertices of different type. The external lines of
such subdiagrams can be either external  or internal with respect to the
whole diagram. To find the contribution of these subdiagrams to the
$S_n$ one has to sum over different ways to attach all lines
external to the subdiagram. This gives the cancelation of divergences
similar to eq-n (\ref{3.16}). The cancelation takes place
even for fixed attachment of external lines to the full diagram $S_n$,
because the $S_n$ is a connected diagram and therefore
there is at least one internal line attached as external to each
subdiagram and the summation over two ways to attach this
line suffices for the cancelation.

Thus we have proved that the sum of all connected
Lorentz-covariant diagrams of order $n$ is UV finite,
and  that for $n\ne 2$ there is no UV divergencies even for fixed attachment of
external lines, while  for $n=2$ one has to have nonzero number of  external
lines and to sum up over different ways of attachment of these external lines.
Only vacuum diagrams of 2nd order and the sums $S_2$ with
fixed attachment of external lines are UV divergent.
Therefore all Green functions without vacuum loops are UV finite.

Let us remark that one cannot prove in similar way the UV 
finiteness of the LF perturbation theory because in this case 
some diagrams become equal to zero. This destroys the described 
cancelation of divergences.

\sect{superpr}{The relation between LF and Lorentz-covariant \st
superpropagators for finite $x^-$}

Let us now consider the boson LF perturbation theory
and compare it with Lorentz-covariant one.
In the LF
formulation the cutoff in LF momentum, $|p_-|\ge\e>0$, is assumed, i.~e.
the Fourier modes of the field $\f(x)$  with $|p_-|<\e$ are excluded. It
can be shown \cite{har,lay}
that the oldfashioned (noncovariant) perturbation theory,
obtained with LF Hamiltonian, can be transformed to equivalent perturbation
theory of Feynman type with the same integrands as in covariant diagrams, but
in corresponding Feynman integrals
the integration over $p_+$ is to be performed
before the integration over $p_-$
and the $p_-$ integration has to be taken over the domain $|p_-|\ge\e>0$.
One can easily derive these facts formally via suitable 
transformations in functional integral \cite{nov88}.
Corresponding Feynman diagrams are called in the following LF diagrams.

In the paper \cite{20}
a method was proposed to find all diagrams (to all
orders of perturbation theory) that give different results
(in the $\e\to 0$ limit) when
calculated in LF and Lorentz-covariant ways.  Having these
diagrams found one can add to canonical LF Hamiltonian counterterms,
compensating the mentioned difference between LF and
Lorentz-covariant   perturbation theories. However  straightforward
application of the method of the paper \cite{20}  to the theory with
Lagrangian (\ref{3.0})
is impossible due to nonpolynomial type of the interaction
term (this  term generates infinite number of diagrams of a given order
in $\g$, and some of partial infinite sums of these diagrams are
UV divergent). We will overcome this difficulty in Sect.~6,
truncating the series for the exponents in the interaction.
After application of the method of the paper \cite{20} we remove
this truncation adding, if necessary, new
counterterms to LF Hamiltonian to make the removing of the
truncation correct.
Let us find, first of all, the difference between the LF and
Lorentz-covariant superpropagators.

Let us consider conventional covariant propagator (\ref{3.4}),
introducing  UV regularization parameter $\Lambda$:
 \disn{4.18}{
\De_{\La}(x)=\frac{i}{\pi}\int\limits_{-\La}^{\La}dk_1\int dk_0
\;\frac{e^{ik_0x^0+ik_1x^1}}{k_0^2-k_1^2-m^2+i0},
\nom}
It can be rewritten in the following form (after integration over $k_0$):
 \disn{4.19}{
\De_{\La}(x)=\int\limits_{-\La}^{\La}dk_1
\;\frac{e^{ik_1x^1-iE(k_1)|x^0|}}{E(k_1)},\quad E(k_1)\equiv \sqrt{k_1^2+m^2}.
\nom}
Introducing new integration variable
 \disn{4.20}{
k=\frac{1}{\sqrt{2}}\ls E(k_1)+k_1\si(x^0)\rs
\nom}
instead of $k_1$, one gets after simple transformations the following
expression:
 \disn{4.24}{
\De_{\La}(x)=\int\limits_{\e_{\La}}^{V_{\La}}\frac{dk}{k}
\;e^{-i\ls kx^-+\frac{m^2}{2k}x^+\rs\si(x^0)},
\nom}
where
 \disn{4.24.1}{
\e_\La=\frac{1}{\sqrt{2}}\ls E(\La)-\La\rs,\qquad
V_\La=\frac{1}{\sqrt{2}}\ls E(\La)+\La\rs.
\nom}

In the LF formulation the propagator, regularized  by the cutoff
$0<\e\le|p_-|\le V$ has the form:
 \disn{4.35}{
\De_{\e,V}^{lf}(x)=\frac{i}{\pi}
\ls\int\limits_{-V}^{-\e}+\int\limits_{\e}^V\rs dk_-
\int dk_+\;\frac{e^{i(k_+x^++k_-x^-)}}{2k_+k_--m^2+i0},
\nom}
or, if $x^+\ne 0$, after the integration over $k_+$,
 \disn{4.36}{
\De_{\e,V}^{lf}(x)=
\int\limits_{\e}^V\frac{dk}{k}
\;e^{-i\ls kx^-+\frac{m^2}{2k}x^+\rs\si(x^+)}.
\nom}
At $x^2\ne 0$  one can transform further these expressions taking the
limit $\e_{\Lambda},\e \to 0$, $V_{\Lambda}, V \to \infty$
and  changing after that the integration variable $k\to k/|x^-|$:
 \disn{4.25b}{
\De_{\La}(x)\to \De(x)=\int\limits_0^{\infty}\frac{dk}{k}\;
e^{-i\ls k+\frac{m^2}{4k}x^2\rs \si(x^0x^-)},
\nom}
 \disn{4.37}{
\De_{\e,V}^{lf}(x)\to \De^{lf}(x)=\int\limits_0^{\infty}\frac{dk}{k}\;
e^{-i\ls k+\frac{m^2}{4k}x^2\rs \si(x^2)}.
\nom}
At $x^2 >0$ one has $\si(x^0x^-)=\si(x^2)=1$. At $x^2 <0$ one can show that
$\Delta (x)$ and  $\Delta^{lf}(x)$ are real (by looking at these integrals
after the additional replacement of variable, $k\to -m^2x^2/(4k)$),
so that the sign of the argument of the exponent in eq-n (\ref{4.25b})
is irrelevant.
Hence  the $\Delta(x)$ and $\Delta^{lf}(x)$  can be written in the same form:
 \disn{4.25d}{
\De(x)=\De^{lf}(x)=\int\limits_0^{\infty}\frac{dk}{k}\;
e^{-i\ls k+\frac{m^2}{4k}x^2\rs}.
\nom}
Therefore at $x^2\ne 0$ the superpropagators $e^{\pm\Delta(x)}$ and
$e^{\pm\Delta^{lf}(x)}$  also coincide. At $x^2 \to 0$
one has poles or powers in $x^2$
for these superpropagators according to eq-ns (\ref{3.7}),(\ref{3.8}).
We need exact
prescriptions for the poles
of the superpropagators $e^{\Delta(x)}$ and $e^{\Delta^{lf}(x)}$
connecting the vertices of different types. For the superpropagator,
connecting vertices of the same types, we get immediately:
 \disn{4.25d.1}{
e^{-\Delta(x)}=e^{-\Delta^{lf}(x)}.
\nom}

For the Lorentz-covariant propagator we can apply at
$x^2\simeq 0, x^{\mu}\ne 0$ the following form:
 \disn{4.32}{
\De(x)\sim -\ln\ls-\frac{m^2e^{2C}}{4}(x^2-i0)\rs,
\nom}
where $C$ is Euler constant and the factor before the $(x^2-i0)$ is the same
as in the corresponding asymptotic decomposition of McDonald function $K_0$.

Hence we get at $x^2\simeq 0$, $x^{\mu}\ne 0$
 \disn{4.34}{
e^{\De(x)}\sim-\frac{4e^{-2C}}{m^2}\frac{1}{x^2-i0}.
\nom}
One can show that the prescription (\ref{4.34})
is equivalent to the integration of
the $e^{\Delta_{\Lambda}(x)}$ over $x$  at first and then taking the
$\Lambda \to \infty$ limit.
The condition $x^{\mu}\ne 0$ is connected with  the nonintegrability of the
$e^{\Delta(x)}$ at $x^{\mu}=0$.
However the cutoff $\La$ 
plays, in fact, only intermediate role because all UV divergencies
cancel in Green functions (as was shown in Sect.~3). 

For the LF superpropagator we consider the domains $x^- \to 0, x^+\ne 0$ and
$x^+\to 0, x^-\ne 0$  separately. In the domain  $x^- \to 0, x^+\ne 0$
we compare expressions (\ref{4.24}) and (\ref{4.36})
and notice that these expressions can be
identified at $\e_{\Lambda}, \e \to 0, V_{\Lambda}, V \to \infty$
due to the equality $\si(x^+)=\si(x^0)$  in the considered domain. Therefore
for the LF superpropagator we can use the same prescription (\ref{4.34})
in this domain.

At $x^+ \to 0, x^-\ne 0$ we will consider the superpropagator
 $e^{\Delta^{lf}_{\e,V}(x)}$, integrated over $x^+$  at fixed
$\e, V$, and prove that the result of the integration, taken in the
$\e\to 0, V\to \infty$  limit, can be equivalently achieved via
integration of LF superpropagator over $x^+$ with principle value prescription.

Consider small interval $|x^+|\le \delta$.
Outside this interval one can take for  expressions (\ref{4.24})
and (\ref{4.36}) their
$\e\to 0, V\to \infty$ limit and use the formula (\ref{4.34}).
Inside the
interval $|x^+|\le \delta$  we investigate the integral
 \disn{4.41}{
I_{\de}=\int\limits_{-\de}^{\de}dx^+\;e^{\De^{lf}_{\e,V}(x)}
\nom}
in the limit $\e\to 0, V\to \infty$. We have
 \disn{4.44}{
I_{\de}=\int\limits_0^{\de}dx^+\ls e^{\De^{lf}_{\e,V}(x^+,x^-)}+
e^{\De^{lf}_{\e,V}(-x^+,x^-)}\rs=\no
=\int\limits_0^{\de}dx^+\;e^{\De^{lf}_{\e,V}(x)}
\ls 1+e^{\De^{lf}_{\e,V}(-x^+,x^-)-\De^{lf}_{\e,V}(x^+,x^-)}\rs,
\nom}
where according to eq-n (\ref{4.36})
 \disn{4.42}{
\De^{lf}_{\e,V}(-x^+,x^-)-\De^{lf}_{\e,V}(x^+,x^-)=
2i\si(x^+)\int\limits_{\e}^V\frac{dk}{k}\;e^{-i\frac{m^2}{2k}|x^+|}
\sin(kx^-)=\no
=2i\si(x^+)\ls\int\limits_{\e}^V\frac{dk}{k}\;\sin(kx^-)+
\int\limits_{\e}^V\frac{dk}{k}\;\sin(kx^-)
\ls e^{-i\frac{m^2}{2k}|x^+|}-1\rs\rs=\no=
i\pi\si(x^2)+D(x)+O(\e)+O(V^{-1}),
\nom}
where
 \disn{4.42.1}{
D(x)=2i\si(x^+)\int\limits_0^\infty\frac{dk}{k}\;\sin(kx^-)
\ls e^{-i\frac{m^2}{2k}|x^+|}-1\rs
\nom}
and the quantities $O(\e)$ and $O(V^{-1})$ do not diverge
at $x^+\to 0$. Here we used the equality
$\int_0^\infty dk\sin(kx^-)/k=\si(x^-)\pi/2$.
So we obtain
 \disn{4.44.1}{
I_{\de}=\int\limits_0^{\de}dx^+\;e^{\De_{\e,V}^{lf}(x)}
\ls 1-e^{D(x)+O(\e)+O(V^{-1})}\rs.
\nom}
One can show, that at $x^+\to 0$ the function
$D(x)$ can be estimated as $|D(x)|\le c\sqrt{x^+}$.
Let us remark that the function $e^{\Delta^{lf}_{\e,V}(x)}$
has a pole in $x^+$ regularized by the parameter $\e$,
and, hence, at $|x^+|\le\e$ this function is bounded by a quantity
of order $\ln(1/\e)$.
Therefore the "divergency"  at lower
limit in the integral (\ref{4.44.1})
is of the order $(O(\e)+O(V^{-1}))\ln\e$,
which is negligible in the $V\to \infty$, $\e\to 0$ limit.

Thus  we get the equality
 \disn{4.45}{
\lim_{\de\to 0}\lim_{\e\to 0}\lim_{V\to\infty}I_{\de}=0,
\nom}
that can be considered as the evidence of the validity of the principle value
prescription in $x^+$  for the LF superpropagator at
$x^+\sim 0$, $x^-\ne 0$.

Hence  the analog of eq-n (\ref{4.34})
for the LF superpropagator should have the form
 \disn{4.47}{
e^{\De^{lf}(x)}\sim -\frac{4e^{-2C}}{m^2} \ls{\cal P}\frac{1}{x^+}\rs
\frac{1}{2x^--\si(x^+)i0},
\nom}
where the ${\cal P}$ denotes the principle value symbol.
This is the limiting value of the superpropagator at
$V\to\infty$, $\e\to 0$. It can not be used directly in the integrands of
Feynman integrals because the superpropagator at finite  $\e$
and large $x^-$ can differ essentially from its limiting value. 
The regularization  $|p_-|\ge \e$  should be removed only after performing
all integrations in Feynman integrals. In contrast with this,
the regularization $|p_-|\le V$ 
plays intermediate role and can be replaced by the prescription
(\ref{4.47}) at
$x^-\approx 0$ (similarly to the situation with the cutoff $\La$
in Lorentz coordinates). This can be possible, in particular,
because at finite $\e$ the superpropagator $e^{\De^{lf}_{\e}(x)}$
is an integrable function at $x^\m=0$ in the limit $V\to\infty$
(we will use the notation 
$\De^{lf}_{\e}(x)=\lim\limits_{V\to\infty}\De^{lf}_{\e,V}(x)$).
The latter  assertion       
can be verified by straightforward analysis of the asymptotic behavior
at most singular points
leading to the asymptotic behavior (\ref{4.34})
in the domain $x^- \approx 0$, $x^+ \ne 0$, whereas
the $x^+ \to 0$ singularity is regulated by the $\e$. This corresponds
to well known fact that  only $\e$-cutoff of
LF momenta is sufficient for UV regularization
of the LF perturbation theory in "longitudinal momenta"
(because the magnitude of the external
$p_-$ momenta sets the upper bound for the internal ones).
We will assume in the following that the cutoff $|p_-|\le V$ 
is removed.

Taking into account that LF and Lorentz-covariant
superpropagators coincide at $x^2 \ne 0$,
and using (\ref{4.47}), we can write for $x^{\m}\ne 0$
 \disn{4.48}{
e^{\De(x)}-e^{\De^{lf}(x)}=
-\frac{4e^{-2C}}{m^2}\frac{1}{2x^--\si(x^+)i0}
\ls\frac{1}{x^+-\si(x^-)i0}-{\cal P}\frac{1}{x^+}\rs=\no=
-\frac{2\pi ie^{-2C}}{m^2|x^-|}\de(x^+).
\nom}

The equalities (\ref{4.25d.1}) and (\ref{4.48}) are valid for the
limiting (at $\e\to 0$) values of the superpropagators
$e^{\pm\De^{lf}_{\e}(x)}$. Therefore, these equalities can not be applied
directly
in LF diagrams, where one must perform all integrations before the limit 
$\e\to 0$ is taken.
Nevertheless one can find in $x$-space a domain, depending on $\e$
and having the following property: if the point $x^{\m}$ stays in 
this domain when $\e$ changes, then the mentioned equalities remain 
true up to terms vanishing in the $\e\to 0 $ limit, even if one 
modifies the equalities by replacing the limiting value 
$\De^{lf}(x)$ with the $\De^{lf}_{\e}(x)$.
At $|x^+|\ge\beta > 0$, we can take the $\e \to 0$  limit in eq-n (\ref{4.36})
if $\e/\beta \to 0$. In this domain the
modified equalities (\ref{4.25d.1}),(\ref{4.48}) are
valid for all $x^-$. At $x^- \ne 0$ and at any $x^+$ we consider bounded domain
$0<\al\le|x^-|\le W$, and $W\e \to 0$ for  $\e \to 0$.
Due to this condition the integral
 \disn{4.49.3}{
\int\limits_{\e}^\infty\frac{dk}{k}\;\sin(kx^-)=
\si(x^-)\int\limits_{\e|x^-|}^\infty\frac{dk}{k}\;\sin k
\nom}
contained in the eq-n (\ref{4.42}) goes in the $\e\to 0$ limit to
$\si(x^-)\pi/2$.
Thus the modified equalities (\ref{4.25d.1}),(\ref{4.48})
are valid in the domain
 \disn{4.49}{
x^{\m}\in\left\{0<\be\le|x^+|\right\}\cup
\left\{0<\al\le|x^-|\le W\right\},\no
\frac{\e}{\be}\str{\e\to 0}0,\qquad W\e\str{\e\to 0}0.
\nom}

\sect{kanon}{The construction of the counterterm generating \st
the difference  between Lorentz-covariant and LF \st
superpropagators in the bounded domain of $x^-$}

Let us consider the action, corresponding to the Lagrangian 
(\ref{3.0}), in  LF coordinates with the cutoff
$|p_-|\ge\e>0$ and with the additional counterterm of the form
 \disn{5.50.0}{
\int d^2xd^2y\; :e^{i\f(x)}e^{-i\f(y)}:u(x-y),
\nom}
where
 \disn{5.50.1}{
u(x)=\tilde u(x^-)\de(x^+).
\nom}
and the $\ti u(x^-)$ is a bounded function, rapidly decreasing
at the infinity.

Analogously to the formula (\ref{A.4})
of the Appendix~1 we get for the $N$-point Green function
 \disn{5.57}{
G_N(y_1,\dots,y_N)=\sum_{k,l,m=0}^{\infty}
\int\ls\prod_{j=1}^{k+l+2m}d^2x_j\rs\frac{(i\la)^k(i\bar\la)^l\,i^m}{k!l!m!}
\times\no\times
\prod_{j=1}^m\ls u(x_{k+j}-x_{k+l+m+j})
e^{-\De^{lf}_{\e}(x_{k+j}-x_{k+l+m+j})}\rs h_{klm},
\nom}
where $\la=\frac{\g}{2}e^{i\te}$, $\bar\la=\frac{\g}{2}e^{-i\te}$, 
 \disn{5.60}{
h_{klm}=\ls\prod_{i,j=1;i<j}^{k+m}e^{-\De^{lf}_{\e}(x_i-x_j)}\rs
\ls\prod_{i,j=1;i<j}^{l+m}e^{-\De^{lf}_{\e}(x_{i+k+m}-x_{j+k+m})}\rs
\times\no\times
\ls\prod_{i=1}^{k+m}\prod_{j=1}^{l+m}e^{\De^{lf}_{\e}(x_i-x_{j+k+m})}\rs
\times\no\times
\langle 0|T
\prod_{i=1}^N\ls\f(y_i)+i\sum_{j=1}^{k+l+2m}
\De^{lf}_{\e}(y_i-x_j)\si(k+m-j+\frac{1}{2})
\rs|0\ra.
\nom}
and $T$ is "LF time" ordering symbol.

Let us remark that $h_{klm}=h_{k+m,l+m,0}$.
Introducing new summation indices  $\ti k=k+m$,
$\ti l=l+m$ we  rewrite
the formulae (\ref{5.57}) in the form:
 \disn{5.62}{
G_N(y_1,\dots,y_N)=\sum_{\ti k,\ti l=0}^{\infty}
\int\ls\prod_{j=1}^{\ti k+\ti l}d^2x_j\rs
\frac{(i\la)^{\ti k}(i\bar\la)^{\ti l}}{\ti k!\ti l!}
\sum_{m=0}^{\min(\ti k,\ti l)}C_{\ti k}^mC_{\ti l}^mm!\times\no\times
\prod_{j=1}^m\ls\frac{-i}{|\la|^2}
u(x_{\ti k-j+1}-x_{\ti k+\ti l-j+1})
e^{-\De^{lf}_{\e}(x_{\ti k-j+1}-x_{\ti k+\ti l-j+1})}\rs h_{\ti k\ti l0},
\nom}
where $C^m_n =\frac{n!}{m!(n-m)!}$.

We transform further this expression, taking into account the locality in time
of the $u(x)$, eq-n (\ref{5.50.1}), and the estimation
$e^{-\De^{lf}_{\e}(x)}\mid_{x^+=0} = O(\e)$ which follows from the eq-n
(\ref{4.36}), so that, if contained in the integrand,
 \disn{5.63}{
u(x_i-x_j)u(x_i-x_k)e^{-\De^{lf}_{\e}(x_j-x_k)}=O(\e).
\nom}
This enables us to rewrite the eq-n (\ref{5.62}) in the form
 \disn{5.65}{
G_N(y_1,\dots,y_N)=\sum_{\ti k,\ti l=0}^{\infty}
\int\ls\prod_{j=1}^{\ti k+\ti l}d^2x_j\rs
\frac{(i\la)^{\ti k}(i\bar\la)^{\ti l}}{\ti k!\ti l!}\times\no\times
\prod_{i=1}^{\ti k}\prod_{j=1}^{\ti l}\ls
1-\frac{i}{|\la|^2}u(x_i-x_{\ti k+j})
e^{-\De^{lf}_{\e}(x_i-x_{\ti k+j})}\rs h_{\ti k\ti l 0}+O(\e),
\nom}
that can be checked using eq-n (\ref{5.63}), the symmetry of the
$h_{\ti k\ti l 0}$  under the permutation of variables
$x_1,\dots,x_{\ti k}$ and of the variables
$x_{\ti k+1},\dots,x_{\ti k+\ti l}$, as well as the possibility to replace
these variables in the integrals of the eq-n (\ref{5.65}).

Let us remark that the eq-n (\ref{5.65}) resembles the expression for the Green
function  (\ref{A.4}) for the theory without counterterms
if the superpropagator $e^{\De^{lf}_{\e}(x)}$ is replaced by
 \disn{5.67}{
e^{\De^{lf}_{\e}(x)}\ls 1-\frac{i}{|\la|^2}u(x)
e^{-\De^{lf}_{\e}(x)}\rs=e^{\De^{lf}_{\e}(x)}-
\frac{4i}{\g^2}\ti u(x^-)\de(x^+).
\nom}
According to eq-n (\ref{4.48}) we can choose
the form of the $\ti u(x^-)$
so that the expression (\ref{5.67}) coincides in the domain (\ref{4.49})
with Lorentz-covariant superpropagator
$e^{\De (x)}$ in the $\e \to 0$ limit. This form is
 \disn{5.69}{
\ti u(x^-)=\frac{\pi}{2}e^{-2C}\frac{\g^2}{m^2|x^-|}
\te(|x^-|-\al)v(\e x^-),
\nom}
where the factor $\te(|x^-|-\al)$ with $\al>0$ is introduced
to regularize the pole in $x^-$. Furthermore for generality an 
arbitrary continuous function $v(\e x^-)$, rapidly decreasing at
infinity and satisfying  the condition  $v(0)=1$, is included.
The decrease of the function $v(\e x^-)$ guarantees that the function 
$\ti u(x^-)$ also decreases as it was required earlier 
(see the text after eq-n (\ref{5.50.1})).
The functions, introduced above, can influence
the function $\ti u(x^-)$ only outside of the domain (\ref{4.49}).

Now we can get the expression (\ref{5.67}) via
addition of the quantity 
 \disn{5.69.1}{
-\frac{2\pi i}{m^2}e^{-2C}\de(x^+)
\frac{\te(|x^-|-\al)}{|x^-|}v(\e x^-)
\nom}
to the LF superpropagator $e^{\De^{lf}_{\e}(x)}$.

Thus the LF theory with the counterterm
 \disn{5.70.2}{
S_c=\frac{\pi}{2}e^{-2C}\frac{\g^2}{m^2}
\int d^2xd^2y\ls:e^{i\f(x)}e^{-i\f(y)}:-1\rs\times\no\times
\de(x^+-y^+)\te(|x^--y^-|-\al)\frac{v(\e(x^--y^-))}{|x^--y^-|}=\no
=\frac{\pi}{2}e^{-2C}\frac{\g^2}{m^2}\int dx^+\biggl(
\int dx^-dy^-\ls :e^{i\f(x^+,x^-)}e^{-i\f(x^+,y^-)}:-1
\rs\times\no\times
\te(|x^--y^-|-\al)\frac{v(\e(x^--y^-))}{|x^--y^-|}\biggr)
\nom}
can be reformulated as a theory with
superpropagator which is equal to a Lorentz-covariant one 
in the domain (\ref{4.49}), but, perhaps, differs from it outside this domain.
In eq-n (\ref{5.70.2}) we subtract the unity from the product of the
exponents (that is irrelevant for physical results, because  one  gets
only additional constant in the action). Besides, we remove the
$\de$-function  integrating over $y^+$. One can see now, that the
addition to the action of local in time (but not in the $x^-$) counterterm $S_c$ 
is equivalent to the addition of corresponding
counterterm to the LF Hamiltonian.
Let us remark also, that the expression (\ref{5.70.2}) must be real.
So we impose the additional condition
$v^*(z)=v(-z)$ on the function $v(z)$, introduced above.

The compensation of the difference between LF and Lorentz-covariant
superpropagators at finite $x^-$ (i.~e. in the bounded domain (\ref{4.49})),
what was done in this and previous sections, is not needed for usual
theories with polynomial interaction, because in such theories
the mentioned difference is absent in the limit $\e\to 0$.
The domain of unbounded $x^-$ plays also an essential role. This 
domain corresponds, in a sense, to the vicinity of the point $p_-=0$.
Therefore, this domain is relevant for the estimation of difference 
between LF and Lorentz-covariant perturbation theory even for the 
theories with polynomial interaction. In the next section we will 
show how to correct the whole LF perturbation theory to make it
equivalent with Lorentz-covariant one, using the compensation
method described above and taking into account the domain of 
unbounded $x^-$.

\sect{sravn}{Complete comparison of LF and Lorentz-covariant \st
perturbation theories}

Consider a perturbation theory for the Green functions without vacuum 
loops, generated by LF Hamiltonian, which is deduced from the action
 \disn{6.122}{
S=\int d^2x\ls\frac{1}{8\pi}\ls \dd_\m\f\dd^\m\f-m^2\f^2\rs+
B:e^{i\f}:+B^*:e^{-i\f}:\rs+\no+
\frac{2\pi}{m^2}e^{-2C}|B|^2\int d^2xd^2y\ls:e^{i\f(x)}e^{-i\f(y)}:-1\rs
\times\no\times
\de(x^+-y^+)\te(|x^--y^-|-\al)\frac{v(\e(x^--y^-))}{|x^--y^-|}
\nom}
by canonical LF quantization. We will show that this
perturbation theory
is equivalent in the limit $\al\to 0$, $\e\to 0$ to usual Lorentz-covariant 
perturbation theory, corresponding to the
Lagrangian (\ref{3.0}). The formula (\ref{6.122})
contains some complex quantity $B$,
which is a function of the parameters $\g$, $\te$, $\al$ and $\e$.
Perturbatively it can be decomposed as follows:
 \disn{6.90}{
B=\frac{\g}{2}e^{i\te}+\sum_{k=2}^\infty B_k\g^k.
\nom}
Notice that in the expression (\ref{6.122}) the term
linear in $B$ coincides in lowest (1st) order
in $\g$ with the interaction term (\ref{3.2}) of the Lagrangian.
The term in the action (\ref{6.122}),
quadratic in $B$, coincides in lowest (2nd) order
with the counterterm (\ref{5.70.2}).

Let us start with diagrams of 2nd order in $\g$ corresponding to
Lagrangian (\ref{3.0})
(Lorentz-covariant and LF results obviously coincide in the 1st the order).
Disconnected diagrams of 2nd order can be considered similar to the 1st
order ones. Connected diagrams can differ by the type of vertices and by  the
type of attachment of external lines. 
Let us at first consider connected diagrams with two vertices of the 
same type. The sum
of all diagrams of this kind with fixed attachment of
external lines is UV finite. If all external lines
are attached only to one vertex
(from now on we call these diagrams "generalized tadpole diagrams",
independently of their order), the diagrams are similar to vacuum ones, and
LF result for these diagrams is zero.
But in Lorentz-covariant theory the corresponding result is nonzero.
Hence, to correct LF Hamiltonian
corresponding to the Lagrangian (\ref{3.0})
one should add a counterterm of second order, generating
the results, $\tilde S_2^{1,1}$ and $\tilde S_2^{2,2}$,
of Lorentz-covariant calculation of the sums of all such diagrams
(with both vertices of the 1st or both vertices of the 2nd type, 
respectively):
 \disn{6.90.1}{
\tilde S_2^{1,1}=-\frac{\g^2}{4}e^{2i\te}
\int d^2x\;\ls e^{-\De(x)}-1\rs,\no
\tilde S_2^{2,2}=-\frac{\g^2}{4}e^{-2i\te}
\int d^2x\;\ls e^{-\De(x)}-1\rs.
\nom}

If the vertices are of the same type and external lines are attached
to both of them, we can write for the sum of these diagrams in Lorentz-covariant
theory the following expression (up to some vertex factors):
 \disn{6.91}{
S_2^-=\g^2\int d^2x\ls e^{-\De(x)}-1\rs e^{ipx},
\nom}
where $p$ is the total external momentum going through these diagrams,
$px=p_+x^++ p_-x^-$.
We assume that the values of external momenta are "nonspecial".
This means that any partial sum of "minus" components of these momenta is not
equal to zero. The coincidence of Green functions at nonspecial values of
external momenta is enough for their physical equivalence, because all special
values of external momenta form a set of zero measure, and a variation of
Green functions on this set does not change physical results. So we will
consider all diagrams only at nonspecial values of external momenta.
Let us rewrite the eq-n (\ref{6.91}) in the form
 \disn{6.92}{
S_2^-=\g^2\int d^2x\lim_{m\to\infty}\sum_{m'=1}^m\frac{1}{m'!}
\ls -\De(x)\rs^{m'}e^{ipx},
\nom}
and change the order of the integration and of the limiting procedure (it is
possible owing to uniform convergence in $m$ of the integral,
what can be proved, using the fact that the limit of the integrand
is a continuous function):
 \disn{6.93}{
S_2^-=\g^2\lim_{m\to\infty}\int d^2x\sum_{m'=1}^m\frac{1}{m'!}
\ls -\De(x)\rs^{m'}e^{ipx}.
\nom}
For any fixed finite $m$  we can relate this quantity with a theory of
polynomial type (with finite number of diagrams in each order) and apply the
results of the paper \cite{20} (see also Appendix~2). So one can show that for
all diagrams with nonspecial values of external momenta,
excepting diagrams which are (or contain as a
subdiagram) generalized tadpole, the
results of LF and Lorentz-covariant perturbation theories coincide in the
$\e\to 0$ limit.
So in the eq-n (\ref{6.93}) (where the exceptional
momenta mentioned above do not
contribute) one can replace Lorentz-covariant propagator with LF one,
taking the limit which removes LF cutoff parameters:
 \disn{6.94}{
S_2^-=\g^2\lim_{m\to\infty}\lim_{\e\to 0}
\int d^2x\sum_{m'=1}^m\frac{1}{m'!}
\ls -\De^{lf}_{\e}(x)\rs^{m'}e^{ipx}.
\nom}
Using the fact that the sum of the series in eq-n (\ref{6.94})
converges in the limit $\e\to 0$ to continuous function
(i.~e. to the $e^{-\De^{lf}(x)}$) one can prove that the
integration over $x$ and the $\e$ limit converge
uniformly with respect to $m$. Hence we can again change the order
of the limiting procedures and of the integration and write:
 \disn{6.95}{
S_2^-=\g^2\lim_{\e\to 0}
\int d^2x\ls e^{-\De^{lf}_{\e}(x)}-1\rs e^{ipx}.
\nom}
Thus Lorentz-covariant and LF results for the considered sum
of diagrams coincide in the limit $\e\to 0$.

For connected Lorentz-covariant diagrams of the 2nd order with
vertices of different type
and with fixed attachment of external lines we
may have UV divergent sums of diagrams. These sums are related with
corresponding superpropagators which have singular behavior at $x^{\mu} \to 0$.
Owing to UV finiteness of
Green functions, proved above in
Sect.~3, we can
introduce any intermediate  UV regularization
for the Lorentz-covariant propagator
to define corresponding superpropagator at $x^{\mu} \to 0$. Here we choose for
convenience the following one:
 \disn{6.96}{
\De_{reg}(x)=
\cases{\De^{lf}_\e(x), & $\{|x^-|\le\al\}\cap\{|x^+|\le\be\}$ \cr
\De(x), & $\{|x^-|>\al\}\cup\{|x^+|>\be\}$ \cr},
\nom}
where $\al$ and $\be$ are cutoff parameters.
As was remarked already (see the end of Sect.~4)
the superpropagator $e^{\De_{\e}^{lf}(x)}$ is integrable function
at $x^{\mu} = 0$ and  fixed $\e>0$. Thus we get the UV
regularization, which can be removed in the $\al\to 0$, $\be\to 0$ limit.

Let us consider connected 2nd order diagrams with all external
lines attached to only one vertex.
These diagrams are generalized tadpoles. The sum of all
these diagrams in Lorentz-covariant theory is described by the expression:
 \disn{6.97}{
\ti S_2^{1,2}=\ti S_2^{2,1}=-\frac{\g^2}{4}
\int d^2x\ls e^{\De_{reg}(x)}-1\rs,
\nom}
where the $\ti S_2^{1,2}$ and $\ti S_2^{2,1}$ are sums
of diagrams with external lines attached to the vertices
of the 1st and of the 2nd type correspondingly. 
Due to the convergence of this integral at large $x$  one can rewrite it as
 \disn{6.98}{
\ti S_2^{1,2}=\ti S_2^{2,1}=-\frac{\g^2}{4}
\int\limits_{-W}^Wdx^-\int dx^+\ls e^{\De_{reg}(x)}-1\rs+\xi,
\nom}
where $\xi$ denotes the quantity vanishing at $W\to\infty$.
For the $x$ in the domain (\ref{4.49}) we
use the formula (\ref{4.48}), where we write $\De^{lf}_{\e}(x)$
instead of  $\De^{lf}(x)$
according to the remark after the formula (\ref{4.48}).
Furthermore we can introduce the factor  $v(\e x^-)$ as in the
formula (\ref{5.69}), because this factor goes to unity in the
$W\e \to 0$ limit. We get in this way
 \disn{6.99}{
\ti S_2^{1,2}=\ti S_2^{2,1}=-\frac{\g^2}{4}\times\no\times
\int\limits_{-W}^Wdx^-\ls\int dx^+\ls e^{\De^{lf}_\e(x)}-1\rs-
2\pi ie^{-2C}\frac{\te(|x^-|-\al)}{m^2|x^-|}v(\e x^-)\rs+\ti\xi,
\nom}
where the definition (\ref{6.96})
of the propagator in the domain $|x^+|\le \be$,
$|x^-|\le \al$ is taken into account.
Here the $\ti\xi$ denotes the quantity vanishing in the $\e\to 0$
limit which is taken at first, and only then the $W\to\infty$ limit.
Let us rewrite eq-n (\ref{6.99})
in the form:
 \disn{6.100}{
\ti S_2^{1,2}=\ti S_2^{2,1}=-\frac{\g^2}{4}
\!\int\! d^2x\ls e^{\De^{lf}_\e(x)}-1\rs+
\frac{i\pi\g^2 e^{-2C}}{2m^2}
\!\int\! dx^-\frac{\te(|x^-|-\al)}{|x^-|}v(\e x^-)+\hskip -11pt\no+
\frac{\g^2}{4}\int\limits_{|x^-|>\e W}
dx^-\ls\int dx^+\ls e^{\De^{lf}_1(x)}-1\rs-
2\pi ie^{-2C}\frac{v(x^-)}{m^2|x^-|}\rs+\ti\xi,
\nom}
where $\De^{lf}_1(x)\equiv \De^{lf}_{\e}(x)\!\mid_{\e=1}$.
Here in the third term we have rechanged the integration variables $x^-\to x^-/\e$,
$x^+\to \e x^+$.
The first term in this formula equals to zero, because
it is LF generalized tadpole diagram
(this corresponds to known fact that such LF diagrams as well as
vacuum LF diagrams
are equal to zero when the cutoff $|p_-|\ge\e>0$ is assumed).
The second term can be generated by
the counterterm (\ref{5.70.2}) (see Sect.~5),
if one adds corresponding counterterm to canonical LF
Hamiltonian.
Let us assume that we have done this.
The third term can not diverge in the
$W\e\to 0$ limit,
because the variation of this term with respect to variation of $W$  is very
small (two first terms and whole expression $\ti S_2^{1,2}$
do not depend on $W$, and the expression $\ti\xi$
depends weakly on the $W$). Therefore we can write the third term in the
$W\e\to 0$ limit as some finite complex constant:
 \disn{6.100.1}{
\frac{\g^2}{4}\lim_{s\to 0}
\int\limits_{|x^-|>s} dx^-\ls\int dx^+\ls e^{\De^{lf}_1(x)}-1\rs-
2\pi ie^{-2C}\frac{v(x^-)}{m^2|x^-|}\rs.
\nom}
The contribution related with this constant also should be
generated by a new counterterm of second order.
In the following it will be useful to write the expression
(\ref{6.100.1})
(up to a  value vanishing in the $\e\to 0$ limit)
in the form of the difference between the
$\ti S^{1,2}_2$ and the 2nd term
in the eq-n (\ref{6.100}), i.~e. as the sum of corresponding
2nd order Lorentz-covariant generalized tadpoles and of
corresponding contribution generated by the counterterm
(\ref{5.70.2}) but taken with the minus sign. Last term will be
considered as one compensating the direct contribution of
the counterterm (\ref{5.70.2}) to the diagram
of generalized tadpole type in the 2nd order.
This direct contribution
(which is the same for diagrams with all external lines
attached to the vertex of the 1st type and to the vertex of the 2nd
type) can be written as $i\frac{\g^2}{4}w$, where
 \disn{6.100.2}{
w=\frac{2\pi e^{-2C}}{m^2}\int dx^-\frac{\te(|x^-|-\e\al)}{|x^-|}v(x^-).
\nom}

Let us consider now the sum of all connected diagrams of the 2nd order with
vertices of different type and with external lines attached to both vertices.
This sum can be described in Lorentz-covariant theory
as follows (up to some vertex factors):
 \disn{6.101}{
S_2^+=\g^2\int d^2x\ls e^{\De_{reg}(x)}-1\rs e^{ipx}
\nom}
where $p$ is total external momentum going through these diagrams
and $\De_{reg}(x)$ is defined by the eq-n (\ref{6.96}).
Similar to previous case, we get:
 \disn{6.102}{
S_2^+=\int d^2x\ls e^{\De^{lf}_\e(x)}-1\rs e^{ipx}-
2\pi ie^{-2C}\int dx^-\frac{\te(|x^-|-\al)}{m^2|x^-|}v(\e x^-)e^{ip_-x^-}-
\no-
\hskip -5mm
\int\limits_{|x^-|>\e W}
\hskip -5mm
dx^-\!\ls\int\! dx^+\!\ls e^{\De^{lf}_1(x)}-1\rs e^{ip_+x^+}-
2\pi ie^{-2C}\frac{v(x^-)}{m^2|x^-|}\rs e^{ip_-x^-/\e}+\ti\xi.
\nom}
The first term of this expression coincides with LF calculation.
The second term can be generated by the counterterm (\ref{5.70.2}).
The third
term goes to zero in the $\e\to 0$  limit at $p_- \ne 0$ owing to the
convergence of the corresponding integral at the $x^-=0$ (as explained above).
Thus Lorentz-covariant and LF results for the considered sum of diagrams
coincide (for nonexceptional $p_-$).

Thus to make LF and Lorentz-covariant theories perturbatively equivalent to 2nd
order in  $\g$ we have to add to canonical LF Hamiltonian the counterterm
corresponding to the expression (\ref{5.70.2}), and also counterterms
compensating the contribution of the counterterm (\ref{5.70.2}) to
diagrams of generalized tadpole type in the 2nd order,
and furthermore the counterterms, generating the expressions
$\ti S_2^{1,1}$, $\ti S_2^{2,2}$, $\ti S_2^{1,2}$ and $\ti S_2^{2,1}$.
It is easy to show, that the sum of all these counterterms (excepting
the first one) is given by the expression
 \disn{6.103}{
-i\sum_{l=0}^{\infty}\frac{:(i\f)^l:}{l!}
\!\ls \ti S_2^{1,1}+\ti S_2^{1,2}-i\frac{\g^2}{4}w\rs\!=\!
\ls -i\ti S_2^{1,1}-i\ti S_2^{1,2}-\frac{\g^2}{4}w\rs\! :e^{i\f}:
\nom}
(which corresponds the attachment of external lines to the vertex
of the 1st type in the generalized tadpole diagram) and by the expression
 \disn{6.104}{
-i\sum_{l=0}^{\infty}\frac{:(-i\f)^l:}{l!}
\ls \ti S_2^{2,2}+\ti S_2^{2,1}-i\frac{\g^2}{4}w\rs=\no=
\ls -i\ti S_2^{2,2}-i\ti S_2^{2,1}-\frac{\g^2}{4}w\rs :e^{-i\f}:,
\nom}
(which corresponds to the attachment of lines to the vertex of the 2nd type). 
Here we have added nonessential constant at $l=0 $ for convenience.
Using the Wick rotation one can show that the quantities
 \disn{6.104.1}{
\int d^2x \ls e^{\pm\De(x)}-1\rs
\nom}
are imaginary, and, hence, the $\ti S_2^{1,2}$  and the $\ti S_2^{2,1}$
are imaginary, and $\ti S_2^{2,2}=-\hbox{${\tilde S}_2^{1,1}$}^*$.
Let us introduce the notation
 \disn{6.104.2}{
A_2=\frac{1}{\g^2}\ls -i \ti S_2^{1,1}-i\ti S_2^{1,2}\rs=\no=
\frac{i}{4}e^{2i\te}\int d^2x\ls e^{-\De(x)}-1\rs+
\frac{i}{4}\int d^2x \ls e^{\De_{reg}(x)}-1\rs
\nom}
such that the $A_2\g^2$ is the sum of connected Lorentz-covariant
generalized tadpole diagrams of the 2nd order with  external lines,
attached to the vertex of the 1st type  (these diagrams should be calculated
with the help of the regularization (\ref{6.96})). We remark that
 \disn{6.104.3}{
-i\ti S_2^{2,2}-i\ti S_2^{2,1}-\frac{\g^2}{4}w=
\ls -i\ti S_2^{1,1}-i\ti S_2^{1,2}-\frac{\g^2}{4}w\rs^*,
\nom}
and conclude that the action (\ref{6.122}) contains all necessary
counterterms of the 2nd order in  $\g$, if one takes
 \disn{6.104.4}{
B_2=A_2-\frac{1}{4}w.
\nom}
This quantity is finite in the  $\e\to 0$, $\al\to 0$ limit due to
the finiteness of the quantity (\ref{6.100.1}) and of the $\ti S_2^{1,1}$
(despite of the divergency of the $w=w(\e\al)$  at
$\e\al\to 0$).

Thus in the 2nd order in $\g$ the LF perturbation theory, generated by the
action (\ref{6.122})  at some finite $B_2$, coincides in the
$\e\to 0$, $\al\to 0$ limit with the Lorentz-covariant perturbation theory,
generated by the Lagrangian (\ref{3.0}).

Notice, that the coincidence of LF and
Lorentz-covariant  expressions for the sum of all connected diagrams of
second  order in $\g$, which are not generalized tadpoles,
was proved only for such
values of external momentum $p$, going through the diagram, that
$p_-\ne  0$  (see, in particular, after  eq-n  (\ref{6.102})).  It is enough for
the coincidence  of  Green  functions in the second order because the
domain $p_-=0$ is a set of zero measure. But the corresponding coincidence
at any $x$ in coordinate space is not guaranteed in principle.
This remark can be important for the $x$ space analysis of higher order
diagrams, which contain the 2nd order diagrams as subdiagrams.
As we will show in sequel, it
suffices for this analysis that the LF and Lorentz-covariant 
expressions in coordinate representation coincide for 2nd order
subdiagrams which are not generalized tadpole and have vertices of 
different type.  Furthermore, this coincidence is needed only in
the domain where the coordinates $x_1^+$, $x_2^+$ of these vertices
are not equal to the corresponding coordinates of the external 
points (with respect to the subdiagram).
Let us prove that this coincidence has place.

Going from diagrams in $p$-space to ones in $x$-space one has to add
factors, corresponding to propagators of external lines
(instead of exponents, containing external momenta).
After the transition to the $x$-space
all reasonings used after the formula (\ref{6.101}) can be repeated,
with the exception of the estimation of the 3d term in the eq-n (\ref{6.102}),
where the condition $p_-\ne 0$ was used.
The analog of this term in the $x$ space can be written as follows (for
simplicity we consider only two external points, $y_1$ and $y_2$):
 \disn{6.103.1}{
-\int d^2x_1d^2x_2\De^{lf}_{\e}(y_1-x_1)\te(|x_1^--x_2^-|-W)\times\no\times
\ls e^{\De^{lf}_\e(x_1-x_2)}-1-2\pi i\de(x_1^+-x_2^+)e^{-2C}
\frac{v(\e(x_1^--x_2^-))}{m^2|x_1^--x_2^-|}\rs\De^{lf}_{\e}(x_2-y_2).
\nom}
The propagator $\De^{lf}_{\e}(z)$  diminishes as $z^-\to\infty$  at
$z^+ \ne 0$. Hence, if $x_1^+ \ne y_1^+$ and $x_2^+ \ne y_2^+$, at least one
of the factors $\De^{lf}_{\e}(y_1 - x_1)$,
$\De^{lf}_{\e}(x_2 - y_2)$ diminish due to
the finiteness of $y_1^-$ and $y_2^-$ and due to the inequality
$|x_1^- - x_2^-|\ge W$.
Therefore the quantity (\ref{6.103.1}) goes to zero in the
$\e\to 0$, $W \to \infty$ limit.

Let us now prove that in the $\e\to 0$, $\al\to 0$ limit
the LF perturbation theory, generated by the action (\ref{6.122}),
coincides with the Lorentz-covariant perturbation theory,
generated by the Lagrangian (\ref{3.0}), to all orders in $\g$
at some choice of the quantities $B_n$.

We consider the sum $S_n$ of all connected Lorentz-covariant
diagrams of order $n>2$ with
fixed attachment of external lines, i.~e. UV finite quantity,
introduced above in Sect.~3. It is convenient
to represent this quantity as a sum of special classes of diagrams.

The $S_n$ can be described in terms of
"incomplete superpropagators" \linebreak $(e^{\pm\De(x)}-1)$. 
The $S_n$ is a sum of terms, corresponding
to connection or disconnection of each pair of vertices by
incomplete superpropagator. Let us rewrite the incomplete
superpropagators as limits of finite sums $D_{\pm}^m(x)$ (they will be
called "cutoff incomplete superpropagators"):
 \disn{6.78}{
e^{\pm\De(x)}-1=\lim_{m\to\infty}D_{\pm}^{m}(x)=
\lim_{m\to\infty}\sum_{m'=1}^{m}\frac{1}{m'!}\ls \pm\De(x)\rs^{m'}.
\nom}
One can write the $S_n$ symbolically as follows
 \disn{6.79}{
S_n=\int d^2x_1\dots d^2x_n\times\no
\times\lim_{m_1,\dots\to\infty}\sum_d
\ls D_+^{m_1}(x_{k_1}-x_{l_1})\times\dots\times
D_-^{m_2}(x_{k_2}-x_{l_2})\times\dots\rs,
\nom}
where the $\sum_d$ denotes the summation over different ways
to connect or disconnect the pairs of vertices by
incomplete superpropagators. According to the definition of the $S_n$
there are no disconnected diagrams among the terms of this sum, but
there are terms which include the generalized tadpole
subdiagrams. There is also possible
that the $S_n$ is a sum of generalized tadpole diagrams.
The generalized tadpole diagrams will play essential
role in the following analysis.

Let us investigate the $S_n$ in Lorentz-covariant perturbation
theory. One can prove, that the functions
$D_-^m(x)$ are integrable uniformly with respect to $m$,
because they tend to continuous function at $m\to\infty$.
To prove the uniform (w.r. to $m$) integrability of the
$D_+^m (x)$ we can use
the behavior (\ref{4.34}) of the superpropagator at $x^2 \to 0, x^{\mu} \ne 0$,
that allows to remove the integration contours
apart from the points with $x^2 = 0$. Furthermore, we can take into account
the cancellation of
UV divergencies at $x^{\mu}= 0$, as shown in Sect.~3 in the proof of the UV
finiteness. This allows to prove that the integrability is uniform
in $m$ in this case also.
Therefore one can change the order of the integration and the limiting
procedures in the eq-n (\ref{6.79}):
 \disn{6.83}{
S_n=\lim_{m_1,\dots\to\infty}\int d^2x_1\dots d^2x_n\times\no
\times\sum_d
\ls D_+^{m_1}(x_{k_1}-x_{l_1})\times\dots\times
D_-^{m_2}(x_{k_2}-x_{l_2})\times\dots\rs.
\nom}
Let us remark that the $\sum_{d}$ in this equation
can be interchanged with the integration (owing to UV finiteness of any finite
sum of diagrams), but it can not be interchanged with the limits in $m_i$,
because, as was shown in Sect.~3,
only the total sum $\sum_{d}$  is UV finite in this limit (not separate terms).

At finite $m_i$ one can apply the method of the paper \cite{20}
to compare the results
of LF and Lorentz-covariant calculations of these quantities.
One concludes (see also Appendix~2) that the
coincidence takes place in the limit $\e\to 0$
in all cases with the exception of generalized
tadpole diagrams  and diagrams including them as subdiagrams. These generalized
tadpole subdiagrams enter as  factors before the rest part of the diagram,
and therefore we can apply the method of the paper \cite{20} only to these rest
parts. Then we can replace  in the expression (\ref{6.83})  Lorentz-covariant
propagators  with  LF ones if these propagators do not enter in generalized
tadpole subdiagrams. In Lorentz-covariant propagators
contained in generalized tadpole diagrams
it is convenient to introduce the regularization (\ref{6.96})
(let us remind, that LF value
of generalized tadpole diagrams is always zero). Thus we can
rewrite the symbolic expression (\ref{6.83}) in the form
 \disn{6.84}{
S_n=\lim_{m_1,\dots\to\infty}\lim_{\al\to 0}\lim_{\e\to 0}
\int d^2x_1\dots d^2x_n\sum_d \biggl(
D_{\e\,+}^{lf\,m_1}(x_{k_1}-x_{l_1})\times\dots\times\no\times
D_+^{m_2}(x_{k_2}-x_{l_2})\times\dots\times
D_{\e\,-}^{lf\,m_3}(x_{k_3}-x_{l_3})\times\dots\times
D_-^{m_4}(x_{k_4}-x_{l_4})\times\dots \biggr).
\nom}
where the $\e$ is LF cutoff parameter, used for the LF propagators,
and the $\al$ is the parameter of the regularization (\ref{6.96})
of  Lorentz-covariant
propagators, participating in  generalized tadpole subdiagrams
and contained in $D_{\pm}^{m_i}$ in the eq-n (\ref{6.84}).

If we could move the limits in $m_i$ through the limits in $\e$, $\al$
and through the operation of integration,
we could write incomplete superpropagators instead of
cutoff incomplete ones and reproduce the Lorentz-covariant result
for the $S_n$ by means of
the canonical LF Hamiltonian with counterterms
generating the contribution equal to
that of Lorentz-covariant generalized tadpole diagrams. However to do this
we need uniform
convergence in $m_i$ of the integrals and of the limits in $\e$, 
$\al$. This convergence depends on
the behavior of the functions $D_{\e\,\pm}^{lf\,m}(x)$ in the integrand.
Repeating the reasonings given before the formula (\ref{6.83}),
we conclude that
there is a difficulty only with the
$D_{\e\,+}^{lf\,m}(x)$ at $x^+ = 0$, $x^- \ne 0$, because LF propagator is
nonanalytical function in $x^+$ at $x^+ = 0$  (see eq-n (\ref{4.36})), and
therefore
it is not allowed to remove the integration contour apart from this singular
point. However we can take into account that the integrand contains the
products of several cutoff superpropagators.
For any 3 vertices one has at least
one superpropagator corresponding to vertices of the same type.
Therefore  if the $x^+$ coordinates of all these 3
vertices tend to coincide, the  behavior of the integrand  at singular point
becomes integrable, because we have for the superpropagator,
connecting the vertices of the same type, the estimation
 \disn{6.86}{
e^{-\De_\e^{lf}(x)}\sim c_1x^+x^-+c_2\e x^-,\qquad x^+\to 0,
\nom}
what improves the convergence in $x^+$ (let us remark that the $\e$  regularizes 
the singularity at $x^+ = 0$ and that it is expected that the domain $x^- \to \infty$ does
not play essential role). This allows to prove
that the contribution of considered integration domain vanishes.
If the $x^+$ coordinates of
two vertices of different type tend to coincide and are not equal to $x^+$
coordinate of some 3d vertex, we can not prove the uniform convergence of the
integrals in this domain. In general case the integration domain can be
subdivided into domains of two classes.
We define these classes in such a way that
in the first class the proof of the uniform convergence fails (i.~e.
the $x^+$ coordinates of some of pairs of vertices of different type
coincide, but are not equal to the $x^+$ coordinates of other vertices), and in
the second class the uniform convergence can be proved. For the second class we
can come back to incomplete superpropagators
from cutoff incomplete ones in the integrand of the
eq-n (\ref{6.84}). For the first class we do the following:
(1) we replace in reverse order the cutoff incomplete LF superpropagators by
corresponding Lorentz-covariant ones (justifying this step as above by
referring to the paper \cite{20});
(2) then we use the uniform convergence property
proved for Lorentz-covariant superpropagators (see the transition from the
eq-n (\ref{6.79}) to the eq-n (\ref{6.83}))
to come back to Lorentz-covariant incomplete
superpropagators from cutoff ones;
(3) after these replacements we remark that the pair
of considered vertices, connected by
Lorentz-covariant incomplete superpropagator,
represents the sum of Lorentz-covariant subdiagrams of the 2nd order, which
can be replaced (as shown above, before the eq-n (\ref{6.103.1})) by the sum of
corresponding LF subdiagrams plus the "vertex"
generated by the counterterm (\ref{5.70.2}),
because, by the definition of the first class, the
$x^+$ coordinates of the
vertices in that pair do not coincide with the $x^+$ coordinates of 
other vertex points. Of course, the generalized tadpole
diagrams, which can be attached to any of the vertices and are collected as
factors before the considered sums of subdiagrams, should be taken into account
separately.

Thus we get
 \disn{6.86.1}{
S_n=\lim_{\al\to 0}\lim_{\e\to 0}\hat S_n,
\nom}
where the $\hat S_n$  is the sum of connected LF diagrams
(as defined at the beginning of Sect.~4) generated by the Lagrangian
(\ref{3.0}) (they are, of course, not generalized tadpoles and do not contain
generalized tadpole subdiagrams) with following modifications: firstly,
to each superpropagator connecting the different type vertices
of these diagrams the contribution, caused by the inclusion of the counterterm 
(\ref{5.70.2}) to the action (see Sect.~5), is added, and, secondly,
all generalized tadpole diagrams from Lorentz-covariant perturbation
theory are added.

We remind also that the limit $\al\to 0$
should be taken only after summing all contributions to Green 
functions  when this limit becomes finite.

Let us show that the sum of connected LF diagrams of
the $n$th order, generated by the action (\ref{6.122}),
coincides with the quantity $\hat S_n$ at some choice of the $B_n$. 
Consider LF perturbation theory  in the parameter
$|B|$ for the action (\ref{6.122}).  We obtain the same set of diagrams,
as with the Lagrangian (\ref{3.0}), but
(1) we have the factors $iB$ and $iB^*$
instead of the factors $i\frac{\g}{2} e^{i\te}$ and $i\frac{\g}{2} e^{-i\te}$
in the vertices,
(2) the quantity (\ref{5.69.1}) is added to every propagator
(as a result of taking into account the term quadratic in $B$ in the action). 
Owing to this addition some connected generalized tadpole diagrams, having 
the order, higher, than the 1st,
and equal to zero in usual LF perturbation theory, may
become nonzero. One can show that only one
generalized tadpole diagram of the 2nd order in $|B|$ is nonzero, 
and it is equal to
 \disn{6.86.2}{
i|B|^2w
\nom}
(that follows strightforwardly from the (\ref{5.69.1})  and from
taking into account the vertex factors  $iB$ and $iB^*$).
These diagrams will be attached to vertices of both types
(to those which have factors $iB$ and $iB^*$), but no more than one
diagram to each vertex. To get this result one has to proceed
analogously to  the  proof of the fact that all
usual connected LF generalized tadpole diagrams (higher than of 1st order)
are equal to zero (see, for example, \cite{20}), and take into account that
the contributions resulting from simultaneous addition of the quantity  
(\ref{5.69.1}) to several propagators, outgoing from one point, 
always equals to zero  (see the eq-n (\ref{5.63})).

Thus starting from the action (\ref{6.122}) we can get all cases,
when generalized tadpole subdiagrams arise, in the following way: one has
to attach to any vertex the generalized tadpole diagram of the 2nd order
in $|B|$ that equals to the quantity (\ref{6.86.2}) (this 2nd order includes
the vertex factor of the mentioned vertex).
If we consider the sum of all contributions, corresponding to a diagram with 
such attachments to some of its  vertices and without this attachments, we
see that both contributions can be described formally as only one, without
the attachments, but with vertex factors $iB$ and $iB^*$ replaced by the
 \disn{6.86.3}{
iB+i|B|^2w,\quad and \quad iB^*+i|B|^2w.
\nom}
Let us now choose the $B$ so that
 \disn{6.86.4}{
B+|B|^2w=A,
\nom}
where the $iA$ is the sum of connected generalized tadpole diagrams 
with all external lines attached to the vertices of the 1st type, generated by the
Lagrangian (\ref{3.0}) in Lorentz-covariant approach.
Then the result of the extraction of the $n$th order
in $\g$ from the set of connected diagrams of lastly considered perturbation
theory with the action (\ref{6.122}) (the $B$ is the series
(\ref{6.90}) in $\g$) coincides with $\hat S_n$, and, hence, in the
limit $\e \to 0$, $\al \to 0$ is equal to the sum of Lorentz-covariant
connected diagrams $S_n$ of order $n$ generated by the Lagrangian 
(\ref{3.0}) (according to the eq-n (\ref{6.86.1})).

Thus we have shown that LF perturbation theory for  connected  Green
functions, generated by the action (\ref{6.122}),   is equivalent 
in the $\e \to 0$, $\al \to 0$ limit to Lorentz-covariant
perturbation theory, corresponding to the Lagrangian (\ref{3.0}).
The coincidence of connected Green functions leads to the coincidence of Green
functions without vacuum loops. 
The value of the quantity $B$, necessary for this coincidence, is given by the
solution of the equation (\ref{6.86.4}).
This equation is fulfilled trivially in the 1st order in  $\g$
($\g B_1=\g A_1=\frac{\g}{2}e^{i\theta}$), while in the
2nd order it coincides with the equation (\ref{6.104.4}),
obtained via straightforward analysis of the 2nd order diagrams. 
As it was shown above (see after the formulae (\ref{6.104.4})),
the quantity $B_2$ is finite.
The consideration of the eq-n (\ref{6.86.4})  in the next orders shows that
the $B_k$  with  $k>2$ are divergent at  removed regularization
(namely, in the limit $\e\al\to 0$).

The eq-n (\ref{6.86.4})  can be solved straightforwardly without using  the
perturbation theory.
We get  the following result:
 \disn{6.86.5}{
B=-\frac{1}{2w}+\sqrt{\frac{1}{4w^2}+\frac{A'}{w}-A''^2}+iA'',
\nom}
where  $A=A'+iA''$, whereas the quantities $A'$, $A''$ are real. 
The sign before the root is fixed by the requirement of the correspondence
with the perturbation theory in $\g$.
Let us remark that at large $\g$ (i.~e. at large fermion mass $M$)
the argument of the root in the eq-n (\ref{6.86.5}) 
may become negative, so that the eq-n (\ref{6.86.4})
can not be  solved. That's why our perturbative 
approach may be applicable only for sufficiently small $M$.

We know that the quantity $iA$ (which
is the sum of connected Lorentz-covariant generalized tadpole diagrams,
generated by the Lagrangian (\ref{3.0}), with the external lines attached
to the vertex of the 1st type) diverges in the $\e\al\to 0$ limit only
in the 2nd order in $\g$ (see Sect.~3). According to eq-n (\ref{6.104.4}) 
we have
 \disn{6.86.6}{
A=\frac{\g^2}{4}w+const
\nom}
(let us remind  that $w\to\infty$ at $\e\al\to 0$). 
Here the $const$ represent a series in $\g$ with finite
(in $\e\to 0$, $\al\to 0$ limit) terms 
starting from $\g A_1$.
Using the eq-n (\ref{6.86.6}), one can take 
the limit $\e\al\to 0$ (and, therefore, the $w\to \infty$ limit) in the eq-n  
(\ref{6.86.5}), whereas this can be done only beyond the perturbation theory. 
We get: 
 \disn{6.86.7}{
B=\sqrt{\frac{\g^2}{4}-A''^2}+iA''=\frac{\g}{2}e^{i\hat\te}, \qquad
\sin\hat\te=\frac{2A''}{\g}.
\nom}
As it is seen from the eq-n (\ref{6.86.6}),  the quantity
$A''$ is finite at $\e\al\to 0$. Remark that the formula
(\ref{6.86.7})  can not be used in the action (\ref{6.122}), 
because the limit $\al\to 0$  can not be taken due to the divergency of
the term quadratic in the $B$.
Therefore one should preserve the regularization in this action
and use there, instead of the eq-n (\ref{6.86.7}),
the expression (\ref{6.86.5}), where the
$B$ is the function of the initial parameters of the theory, $\g$ and $\te$, 
and also of the regularization factor $\e\al$. The quantity
$B$ depends (through the quantity $w$) also on the function $v(z)$,
which is not fixed completely, so that some arbitrariness remains.   
The LF Hamiltonian, corresponding to the action (\ref{6.122}), has the form: 
 \disn{6.123}{
H=\int dx^-\ls\frac{1}{8\pi}m^2:\f^2:
-B:e^{i\f}:-B^*:e^{-i\f}:\rs-
2\pi e^{-2C}\frac{|B|^2}{m^2}\times\no\times
\int\! dx^-dy^-\!
\ls :e^{i\f(x^-)}e^{-i\f(y^-)}:-1\rs
\te(|x^--y^-|-\al)
\frac{v(\e(x^--y^-))}{|x^--y^-|},
\nom}
where the LF cutoffs $|p_-| \ge\e>0$ are implied, and $\al >0$. If the 
function $v(x)$ fulfills the introduced requirements and if the 
coefficient $B$, dependent on $v(x)$, is defined by the eq-n (\ref{6.86.5}), then
the LF Hamiltonian (\ref{6.123}) generates perturbation theory which is equivalent
in the limit $\e\to 0$, $\al\to 0$ 
to  Lorentz-covariant perturbation theory corresponding to the Lagrangian (\ref{3.0}).

\sect{vozvr}{The transformation of boson variables in the corrected \st
LF Hamiltonian to canonical LF fermion variables}

In this section we introduce LF fermionic fields, giving their construction in
terms of boson field variables, i.~e. we make the transformation inverse to the
bosonization. In analogy with the procedure of Sect.~2 we
consider the theory on finite interval $-L\le x^- \le L$  assuming periodic
boundary conditions for the boson field $\f(x)=\sqrt{4\pi}\Phi(x)$.
We replace the cutoff $|p_-|\ge \e >0$ by simple exclusion of zero
mode in the Fourier series
 \disn{7.126}{
\f(x^-)=\sum_{n\ne 0}\f_ne^{i\frac{\pi}{L}nx^-}.
\nom}
The Hamiltonian  (\ref{6.123}) takes the form
 \disn{7.124}{
H=\int\limits_{-L}^Ldx^-\ls\frac{1}{8\pi}m^2:\f^2:
-B:e^{i\f}:-B^*:e^{-i\f}:\rs-
2\pi e^{-2C}\frac{|B|^2}{m^2}\times\no\times
\int\limits_{-L}^Ldx^-\int\limits_{-L}^Ldy^-
\ls:e^{i\f(x^-)}e^{-i\f(y^-)}:-1\rs
\ti\te(x^--y^-,\al)
\frac{v(\frac{x^--y^-}{L})}{|x^--y^-|}.
\nom}
Here the $\ti\te(z,\al)$  is
periodic analog of the function $\te(|z|-\al)$, 
i. e.  $\ti\te(z,\al)=0$, if $2Ln-\al<z<2Ln+\al$ for some integer $n$,
and  otherwise $\ti\te(z,\al)=1$.
The function $v(z)$ also must be limited by the condition of the periodicity of the
integrand (what means the translation invariance of the Hamiltonian in $x^-$)
in the following way
 \disn{7.126.1}{
v(z)=|z|\ti v(z),
\nom}
where the function $\ti v(z)$ is periodic with the period equal to $2$.
Below we use the formulae similar to  (\ref{2.32}) to construct
the LF version of fermion field in terms of boson field variables. With this
aim we rewrite the normal ordered expression of the eq-n
(\ref{7.124}) in the form:
 \disn{7.128}{
:e^{i\f(x^-)}e^{-i\f(y^-)}:=:e^{i\f(x^-)}::e^{-i\f(y^-)}:e^{-\be(x^--y^-)},
\nom}
where
 \disn{7.127}{
\be(x^--y^-)=\sum_{n=1}^\infty\frac{1}{n}e^{-i\frac{\pi}{L}n(x^--y^-)}.
\nom}

The construction, similar to  the formulae  (\ref{2.32}),
should represent only one fermionic field
component on the LF (which remains independent after solving LF
canonical constraints). Choosing the antiperiodic boundary conditions for this
fermionic field  we write the following expression satisfying canonical
anticommutation relations on the LF as a consequence of ones for boson 
variables:
 \disn{7.p1}{
\hat\psi_+(x)=\frac{1}{\sqrt{2L}}e^{-i\hat\om}e^{-i\frac{\pi}{L}x^-\hat Q}
e^{i\frac{\pi}{2L}x^-}:e^{-i\f(x)}:.
\nom}
Here the $\hat \psi_+$, $\hat Q$, $\hat\om$ denote the LF analog of
the quantities $\psi_+$, $Q_+$, $\om_+$, appearing in the
bosonization procedure in Lorentz coordinates (described in Sect.~2),
with the analogous commutation relations, but on the LF.
We also fix the  irrelevant now variable $A_-$ to be equal to zero.

The described construction can be used to express  the boson Hamiltonian
in terms of fermion field on the LF. This expression takes a form
similar to the naive canonical LF $QED_2$
Hamiltonian (when antiperiodic boundary conditions for fermion
field and the condition $A_-=0$ are chosen), if one defines the 
$\ti v(z)$ as follows:
 \disn{7.133}{
\ti v(z)=\exp(\sum_{m=1}^\infty\frac{1}{m}e^{-i\pi mz})
\sum_{n=-\infty}^\infty\frac{1}{n+\frac{1}{2}}e^{i\pi nz}.
\nom}
This function satisfies all necessary conditions. The condition,
required at $z=0$, can
be checked using the following asymptotic forms:
 \disn{7.134}{
\exp(\sum_{m=1}^\infty\frac{1}{m}e^{-i\pi mz})=
\frac{1}{1-e^{-i\pi z}}=\frac{1}{i\pi z}+u_1(z),
\nom}
 \disn{7.135}{
\sum_{n=-\infty}^\infty\frac{1}{n+\frac{1}{2}}e^{i\pi nz}=
2e^{-i\frac{\pi}{2}z}\ls
\ln\ls 1-e^{-i\frac{\pi}{2}z}\rs-\ln\ls 1-e^{i\frac{\pi}{2}z}\rs\rs+\no
+\ln\ls 1-e^{i\pi z}\rs-e^{-i\pi z}\ln\ls 1-e^{-i\pi z}\rs=
i\pi\si(z)+u_2(z),
\nom}
where  $u_1 (z)$, $u_2(z)$ are continuous functions at $z=0$  and $u_2(0) =0$.
Hence the function $v(z)= |z|\ti v(z)$ is continuous and equal
to unity at $z = 0$.
As remarked at the end of the preceding section this is sufficient for having 
 perturbation theory equivalent to the Lorentz-covariant one in 
the limit $\e\to 0$, $\al\to 0$.

The Hamiltonian (\ref{7.124}) can be rewritten as follows:
 \disn{7.136}{
H=\int\limits_{-L}^Ldx^-\ls\frac{1}{8\pi}m^2:\f^2:
-B:e^{i\f}:-B^*:e^{-i\f}:\rs-
2\pi e^{-2C}\frac{|B|^2}{m^2L}\times\no\times
\int\limits_{-L}^L\!dx^-\!\int\limits_{-L}^L\!dy^-
:e^{i\f(x^-)}:\;:e^{-i\f(y^-)}:
\sum_{n=-\infty}^\infty
\frac{\ti\te(x^--y^-,\al)}{n+\frac{1}{2}}\:e^{i\frac{\pi}{L}n(x^--y^-)}.
\nom}
Returning to fermion variables
according to the formulae (\ref{7.p1}) one gets:
 \disn{7.p2}{
H=\int\limits_{-L}^Ldx^-\biggl(\frac{e^2}{2}\ls \dd_-^{-1}
[\hat\psi_+^+\hat\psi_+]\rs^2-
\sqrt{2L}\ls B^* e^{-i\frac{\pi}{2L}x^-}
e^{i\frac{\pi}{L}x^-\hat Q}e^{i\hat\om}\hat\psi_++h.c.\rs\biggr)-\no
-4\pi e^{-2C}\frac{|B|^2}{m^2}
\int\limits_{-L}^Ldx^-\int\limits_{-L}^Ldy^-
\ls\sum_{n=-\infty}^\infty\frac{1}{n+\frac{1}{2}}e^{i\frac{\pi}{L}n(x^--y^-)}
\rs\ti\te(x^--y^-,\al)\times\no\times
\hat\psi_+^+(x^-)
e^{-i\hat\om}e^{-i\frac{\pi}{L}x^-\hat Q}e^{i\frac{\pi}{2L}x^-}
e^{-i\frac{\pi}{2L}y^-}
e^{i\frac{\pi}{L}y^-\hat Q}e^{i\hat\om}\hat\psi_+(y^-),
\nom}
where the $[f(x)]$ denotes the quantity $f(x)$
without its zero mode in $x^-$ like in the eq-n (\ref{2.6}).
Using commutation relations between $\hat\psi_+$, $\hat Q$, $\hat\om$,
which are analogous to ones for $\psi_+$, $Q_+$, $\om_+$ in Sect.~2, and
restricting the Hamiltonian to physical subspace with $\hat Q=0$, we get
 \disn{7.p2.1}{
H=\int\limits_{-L}^Ldx^-\biggl(\frac{e^2}{2}\ls \dd_-^{-1}
[\hat\psi_+^+\hat\psi_+]\rs^2-
\sqrt{2L}\ls B^* e^{-i\frac{\pi}{2L}x^-}
e^{i\hat\om}\hat\psi_++h.c.\rs\biggr)-\no
-4\pi e^{-2C}\frac{|B|^2}{m^2}
\int\limits_{-L}^Ldx^-\int\limits_{-L}^Ldy^-
\ls\sum_{n=-\infty}^\infty\frac{1}{n+\frac{1}{2}}
e^{i\frac{\pi}{L}(n+\frac{1}{2})(x^--y^-)}
\rs\ti\te(x^--y^-,\al)\times\no\times
\hat\psi_+^+(x^-)\hat\psi_+(y^-),
\nom}
It is easy to show that the multiplication of the quantity
(\ref{7.135}) in the formulae (\ref{7.p2.1}) by the function 
$\ti\te(x^--y^-,\al)$ can (at small $\al$) modify noticeably the Fourier 
modes 
of this quantity only at $n \sim\frac{L}{\al}$. However, since the
calculations with LF Hamiltonian are always made at some finite value 
of the $p_-$, this modification can not change the results.
This means that one can take the limit $\al\to 0$ in the Hamiltonian 
(\ref{7.p2.1})
(let us remark that after this one can not develop the perturbation 
theory in $\g$ (i.~e. in fermion mass), because this means actually
the returning to bosons where one has divergency at $\al=0$).
We conclude that one can use, in particular, the
finite expression (\ref{6.86.7}) for the quantity
$B$ instead of the expression  (\ref{6.86.5}).
Then one can rewrite the Hamiltonian
(\ref{7.p2.1})  as follows (using eq-n (\ref{3.2})):
 \disn{7.140}{
H=\int\limits_{-L}^Ldx^-\biggl(\frac{e^2}{2}\ls \dd_-^{-1}
[\hat\psi_+^+\hat\psi_+]\rs^2-\no-
\frac{eMe^C\sqrt{2L}}{4\pi^{3/2}}\ls e^{-i\hat\te-i\frac{\pi}{2L}x^-}
e^{i\hat\om}\hat\psi_++h.c.\rs-
\frac{iM^2}{2}\hat\psi_+^+\dd_-^{-1}\hat\psi_+\biggr).
\nom}
The quantity  $\hat\te$, determined by the eq-n
(\ref{6.86.7}), can be interpreted as some condensate angle.
Indeed, one can show (as in the paper \cite{30}) that
 \disn{7.140.1}{
A=\frac{1}{2}\ls\g\frac{\dd}{\dd\g}+\frac{1}{i}
\frac{\dd}{\dd\te}\rs\ti G_0,
\nom} 
where the $\ti G_0$ is the connected vacuum Green function density:
 \disn{7.140.2}{
\ti G_0=\frac{1}{V}\ln\langle 0|T\exp \ls i\int d^2xL_I \rs |0 \ra.
\nom}
Here the $V$  is the volume of the space-time,
and the $L_I$ is given by the formula
(\ref{3.2}). Therefore one can write in Heisenberg representation:
 \disn{7.140.3}{
A=\frac{\g}{2}\langle \Om|:e^{i(\f+\te)}:|\Om\ra,
\nom} 
\disn{7.140.4}{
\sin\hat\te =\langle\Om|:\sin(\f+\te):|\Om\ra=
-\frac{2\pi^{3/2}}{e\;e^C}\langle\Om|:\bar\Psi\g^5\Psi:|\Om\ra,
\nom} 
where the $|\Om\ra$  is the physical vacuum,
and the normal ordering is taken with
respect to the bare vacuum of Lorentz-covariant perturbation theory. 
The quantity $\hat\te$ is a finite function of the parameters
$M/e$ and $\te$ of initial theory (see the text after the formulae 
(\ref{6.86.7})).
According to the first form of the formulae (\ref{7.140.4}) 
 the quantity $\hat\te$
coincides with the $\te$ in lowest order in $\g$  (i.~e. at $M=0$).
If the fermion mass $M$ (actually $M/e$) increases, the right hand side
of the eq-n (\ref{7.140.4}) may become greater than unity. Then this eq-n
cannot be resolved with respect to $\hat\te$, and our method of
the construction of LF Hamiltonian, using the perturbation theory in $M$,
becomes not working. This was already remarked after the eq-n (\ref{6.86.5}).
Only numerical nonperturbative calculations
can tell us something  more concrete about the appearance of such difficulty.

Let us remark that the last term in the formulae (\ref{7.140})
coincides with the term, which usually appears in direct quantization
on the LF (after solving canonical constraints for fermion fields).

If one writes the expression for $\hat\psi(x)_+$ in terms of canonical fermion
creation and annihilation operators at $x^+=0$:
 \disn{7.p3}{
\hat\psi_+(x)=\frac{1}{\sqrt{2L}}\ls
\sum_{n\ge 1}b_ne^{-i\frac{\pi}{L}(n-\frac{1}{2})x^-}+
\sum_{n\ge 0}d_n^+ e^{i\frac{\pi}{L}(n+\frac{1}{2})x^-}\rs,
\nom}
so that
 \disn{7.p3.1}{
\{b_n,b_{n'}^+\}=\{d_n,d_{n'}^+\}=\de_{nn'},\quad
b_n|0\ra=d_n|0\ra=0,
\nom}
and the momentum $P_-$ and the charge $\hat Q$  take the form
 \disn{7.p4}{
P_- =\sum_{n\ge 1}b_n^+b_n\frac{\pi}{L}(n-\frac{1}{2})+
\sum_{n\ge 0}d_n^+d_n\frac{\pi}{L}(n+\frac{1}{2}),
\nom}
 \disn{7.p5}{
\hat Q=\sum_{n\ge 1}b_n^+b_n-\sum_{n\ge 0}d_n^+d_n,
\nom}
one can express the 2nd term in the eq-n (\ref{7.140})
using only fermionic "zero" modes (after the integration over $x^-$):
 \disn{7.p6}{
P_+=H=\int\limits_{-L}^Ldx^-\biggl(\frac{e^2}{2}\ls \dd_-^{-1}
[\hat\psi_+^+\hat\psi_+]\rs^2-\hskip 40mm\no\hskip 40mm
-\frac{M}{2}\ls R\:e^{i\hat\om}d_0^++h.c.\rs-
\frac{iM^2}{2}\hat\psi_+^+\dd_-^{-1}\hat\psi_+\biggr),
\nom}
where
 \disn{7.p6.1}{
R=\frac{e\:e^C}{2\pi^{3/2}} e^{-i\hat\te}.
\nom}
The complex parameter $R$ can be related with fermion condensates as follows:
 \disn{7.p6.2}{
|R|=\left|\langle\Om|:\bar\Psi\ls 1+i\g^5\rs\Psi:|\Om\ra\right|
\Bigr|_{M=0},\no
{\rm Im}\:R=\langle\Om|:\bar\Psi\g^5\Psi:|\Om\ra.
\nom}

In contrast to our previous paper \cite{30} only finite condensates enter our 
LF Hamiltonian explicitly. This also agrees with general idea of the paper 
\cite{37} that only finite nonperturbative (with respect to usual coupling)
parts of the condensates should play role of physically meaningful parameters.
One can use  the $M$ and the $\hat\te$ as independent parameters in 
nonperturbative calculations with our LF Hamiltonian and fit these parameters
to reproduce known results.

Let us remark that the phase operator $e ^{i\hat\om}$, present in the LF
Hamiltonian (\ref{7.p6}),  is defined  on the LF  analogously
to formulas (\ref{2.29}-\ref{2.31}),
with the LF analog of "filled" states (\ref{2.15}).  This means that we have
 \disn{7.p6.3}{
[\hat\om,\hat Q] = i, \qquad
e^{i\hat\om}\hat Qe^{-i\hat\om}= \hat Q - 1,
\nom}
and therefore, using the formulae (\ref{7.p1}), we can write
 \disn{7.p6.4}{
e^{i\hat\om}\hat\psi_+(x)e^{-i\hat\om} 
=e^{i\frac{\pi}{L}x^-}\hat\psi_+(x),               
\nom}
so that 
 \disn{7.p6.5}{
e^{i\hat\om}b_n e^{-i\hat\om}=b_{n+1},\quad
e^{i\hat\om} d_n^+ e^{-i\hat\om}=d_{n-1}^+,\qquad  n\ge 1,\no
e^{i\hat\om} d_0^+ e^{-i\hat\om} =b_1.
\nom}
Besides we have
 \disn{7.p6.6}{
e^{i\hat\om}|0\ra=b^+_1 |0\ra,\qquad
e^{-i\hat\om}|0\ra=d^+_0 |0\ra,
\nom}
because the vacuum $|0\ra$  has the following  "Dirac see" form on the LF:
 \disn{7.p6.7}{
|0\ra = \ls \prod_{n=0}^{\infty} d_n\rs |0_D\ra.
\nom}

These equalities help to calculate matrix elements
of the Hamiltonian (\ref{7.p6}) on
physical subspace, defined by the condition
 \disn{7.p6.8}{
\hat Q |phys\ra=0.
\nom}

As simplest example one can consider $q\bar q$ approximation
for  boson bound state wave functions and compare the results
of the calculation of the spectrum with known ones.

It is more interesting, however,  to find a way of generalization of
our 2-di\-men\-si\-o\-nal
analysis to 4-dimensional $QCD$. Such possibility  can be connected with the 
assumption that at small $M$ nonperturbative contributions related with 
condensates  enter LF Hamiltonian by means of only zero modes of
quark and gluon fields with the coefficients depending on condensate 
parameters in analogy with  considered here LF $QED_2$. 
This assumption was already used in our earlier paper \cite{14}. Similar idea
was discussed  also in  \cite{mac1,mac2}.

\sect{zakl}{Conclusion}

In conclusion let us formulate the results. We considered
the problem of finding LF Hamiltonian which gives the theory
equivalent to conventional one (in Lorentz-covariant
formulation). For the models with gauge symmetry this
problem usually is very difficult one. In the present paper
we solved it successfully for QED in (1+1)-dimensions. The
following methods were used: (1) we transformed the $QED_2$
to its bosonized version, which is  known theory of
selfinteracting scalar field (with fermion mass $M$ playing
the role of coupling constant), (2) in this bosonized theory
we applied the method proposed earlier to
find the difference between LF and Lorentz-covariant
perturbation theories (in small parameter $M$), what
gives us the counterterm in LF Hamiltonian that  can
compensate this difference, (3) then we made "inverse"
transformation from boson variables to initial fermionic
ones directly on the LF (in fact we have made the transformation
to  only  that
component of fermion field which remains independent after
solving  canonical  LF constraints for fermion fields).
In  the obtained LF Hamiltonian  (\ref{7.p2.1})  we used
the formulation with discretized LF momentum (more convenient for
nonperturbative calculations),
choosing antiperiodic boundary conditions for fermion fields
(while starting with periodic ones for the
boson field) on the  LF interval $|x^-|\le L$.
The UV-type regularization is present in this Hamiltonian,
and the coefficients
before the counterterms are perturbative series
in fermion mass with
terms diverging in the limit of removing the regularization.
However, after the summation of these series,  the mentioned
coefficients turn out to be finite in the limit
of removing the regularization. 
This allowed us to formulate the LF Hamiltonian in the final form
(\ref{7.p6}), where the limiting
values for these coefficients were used. 
In this form the Hamiltonian can not be longer used for the
construction of perturbation theory in $M$. But it can be applied for 
nonperturbative calculations.

This final LF Hamiltonian has, beside of usual terms, obtained via
naive canonical  LF quantization of $QED_2$ in LF gauge, a counterterm,
proportional to linear combination of fermionic zero modes
$(\psi_+^+)_0$ and $(\psi_+)_0$ (multiplied by some operator phase factors
neutralizing their charge and fermionic number), with
coefficients before them ($-MR/2$ and $-MR^*/2$, correspondingly) 
proportional to fermion mass $M$ and depending on fermion condensates:
 \disn{8.1}{
|R|=\left|\langle\Om|:\bar\Psi\ls 1+i\g^5\rs\Psi:|\Om\ra\right|
\Bigr|_{M=0}=\frac{e\:e^C}{2\pi^{3/2}},\no
{\rm Im}\:R=\langle\Om|:\bar\Psi\g^5\Psi:|\Om\ra=-|R|\sin\hat\te.
\nom}

Such final LF Hamiltonian generates a theory equivalent to
Lorentz-co\-va\-ri\-ant
$QED_2$  (in the $L\to\infty$ limit), if the fermion mass $M$
is small enough to 
guarantee the consistency of the eq-ns (\ref{8.1}).

One can use the $M$ and the $\hat \te$ as input parameters in
nonperturbative (numerical) calculations with the obtained
LF Hamiltonian and fit the $\hat \te$ to known spectra.

The form of the obtained counterterm is rather simple and allows to hope 
that for 4-dimensional  $QCD$  LF Hamiltonian the  effects, related with
condensates, can be taken into account with only zero modes
of fields like in the $QED_2$  (at least at small $M$  and
semi-phenomenologically).

\vskip 5mm
{\bf Acknowledgements.}
Part of this work was carried out during a stay of E.~P. at the University of
Erlangen and supported by the Deutsche Forschungemeinschaft (grant DFG
436 RUS 113/324/0(R)). E.P. thanks F.Lenz, M.Thies and B.van de Sande for
useful discussions and help. This work was also supported in part (for S.P.)
by the Gribov Scholarship of the World Federation of Scientists.

\setcounter{form}{0}
\renewcommand{\theform}{{\rm A}1.\arabic{form}}

\section*{$\protect\vphantom{a}$\hfill Appendix 1}

Perturbative form of Green functions in terms of the superpropagators.

The $N$-point Green function $G_N(y_1,\dots,y_N)$ has the following form in the
interaction picture:
 \disn{A.1}{
G_N(y_1,\dots,y_N)=\langle 0|T\ls \f(y_1)\dots
\f(y_N)e^{i\int d^2xL_I}\rs|0\ra,
\nom}
where $T$ is "time" ordering symbol and
 \disn{A.2}{
L_I=\frac{\g}{2}e^{i\te}:e^{i\f}:+\frac{\g}{2}e^{-i\te}:e^{-i\f}:.
\nom}
We expand the exponent in perturbation theory series:
 \disn{A.3}{
G_N(y_1,\dots,y_N)=\sum_{l,m=0}^{\infty}\ls\frac{i\g}{2}\rs^{l+m}
\frac{e^{i(l-m)\te}}{l!m!}
\int\ls\prod_{i=1}^{l+m} d^2x_i\rs
\times\no\times\langle 0|T
\ls\prod_{n=1}^N\f(y_n)\rs\ls\prod_{i=1}^l:e^{i\f(x_i)}:\rs
\ls\prod_{j=1}^m:e^{-i\f(x_{l+j})}:\rs|0\ra.
\nom}
At any fixed order of times in the time-ordered product  we can calculate
corresponding vacuum matrix element by transposing the exponents with only creation
operators to the left and those with annihilation operators  to the right. The
result can be expressed in terms of propagators $\Delta (x)$  as follows
\cite{adam}:
 \disn{A.4}{
G_N(y_1,\dots,y_N)=\sum_{l,m=0}^{\infty}\ls\frac{i\g}{2}\rs^{l+m}
\frac{e^{i(l-m)\te}}{l!m!}
\int\ls\prod_{i=1}^{l+m} d^2x_i\rs\times\no\times
\ls\prod_{i,j=1;i<j}^l e^{-\De(x_i-x_j)}\rs
\ls\prod_{i,j=1;i<j}^m e^{-\De(x_{l+i}-x_{l+j})}\rs
\ls\prod_{i=1}^l\prod_{j=1}^m e^{\De(x_i-x_{l+j})}\rs\times\no\times
\langle 0|T
\prod_{n=1}^N\ls\f(y_n)+i\sum_{j=1}^{l+m}\De(y_n-x_j)\si(l-j+\frac{1}{2})
\rs|0\ra.
\nom}

This formulae shows that perturbation theory series  is  expressed actually
in terms of superpropagators  $e^{\pm\De(x)}$, which connect
"internal"  points having coordinates $x_i$,  and  Lorentz-covariant propagators,
which connect these points with the "external" points, denoted by the
$y_i$.

\setcounter{form}{0}
\renewcommand{\theform}{{\rm A}2.\arabic{form}}

\section*{$\protect\vphantom{a}$\hfill Appendix 2}

Let us describe briefly the method of the paper \cite{20}
allowing to compare  LF and Lorentz-covariant Feynman integrals
in the theory with polynomial interaction. Consider, for
example,  an  arbitrary  1-loop  Feynman  diagram  with  external
entering momenta $p^i_{\mu}$, $i=1,2,\dots$.  The  loop  momentum
$k_-$ is bounded by cutoff  conditions  $|k_--\sum  p^i_-|\ge\e$,
steming from the restriction on the propagator momenta. On the  other  side,
analogous covariant diagram contains  the  integration  over  all
$k$. Therefore the difference between these diagrams can be found
as the sum of integrals over the bands $ |k_- - \sum p_-^i|<\e$.

Let us estimate one of these "$\e$"-band integrals. We shift  the
variable $k_-$ in this  integral  so  that  $|k_-|<\e$.  Then  we
change the scale:
 \disn{B11a}{
k_-\longrightarrow \e k_-,\qquad k_+\longrightarrow \frac{1}{\e} k_+.
\nom}
that makes the integration interval independent of $\e$, while
keeps unchanged Lorentz-invariant products like $k_+k_-$ or $dk_+dk_-$.
Propagators, corresponding to internal lines whose momenta are
outside of the $\e$-band (owing to external momentum $p_-$ going
through the line) change as follows:
 \disn{B11b}{
\frac{i}{\pi}\,\frac{1}{2(k_++p_+)(k_-+p_-)-m^2+i0}\longrightarrow\no
\longrightarrow
\frac{i}{\pi}\,\frac{1}{2(\frac{1}{\e}k_++p_+)(\e k_-+p_-)-m^2+i0}
\mathrel{\mathop{\approx}\limits_{\e\to 0}}
\frac{i}{\pi}\,\frac{\e}{2k_+p_-}.
\nom}
In the paper \cite{20} we used denotations for the  lines  with
momenta outside and inside of the $\e$-band. The first one was called
$\Pi$-line  and  the  last  one  $\e$-line.  It  follows    from the
eq-n(\ref{B11b}) that every $\Pi$-line gives  a  factor  of  order
$O(\e)$ while every $\e$-line gives a  factor  of  order  $O(1)$.
Therefore the integral over the band is zero in $\e\to  0$  limit
if at least one of $\Pi$-lines is present in the corresponding diagram.

Similar analysis can be made for an arbitrary  many-loop  Feynman
diagram. The difference between LF and covariant  calculation  of
this diagram can be estimated again by considering  all  possible
configurations  of  $\Pi$-  and  $\e$-  lines  in  the    diagram
\cite{20}. It was shown in the paper \cite{20}  (for  a  wide
class of field theories) that each of these configurations can be
estimated as having the order $O(\e^{\s})(1+O(\log\e))$
with respect to $\e$, where
 \disn{B11}{
\s=min(\ta,\om_--\om_+-\m+\et).
\nom}
Here the minimum is to be taken w.r.t. all  subdiagrams  of  the
diagram at some configuration of $\Pi$- and $\e$-  lines  in  it;
$\omega_{\pm}$ are indices of UV-divergency  in  $k_{\pm}$  of  a
given subdiagram; $\mu$ is the index of  total  UV-divergency  in
$k_-$ of all $\Pi$-lines in the subdiagram; $\tau$ is the total power
of the $\e$ that arises, after the change $k_-\to \e k_-$ of loop
variables $k_-$,  from  numerators  of  all  propagators  of  the
diagram and from all volume elements in the integrals over $k_-$;
$\eta$ is  the  part  of  the  $\tau$  related  with  only  those
numerators and volume elements (used in the definition of $\tau$)
that are not present in the considered subdiagram. Let  us  apply
this general result to our scalar field theory.  All  propagators
have simple structure. Only possible contribution to  the  $\tau$
comes from the volume elements $dk_-$. Because we are  interested
only  in  the  difference  of  LF  and  covariant  diagrams,  any
configuration should contain at least one integration over  $k_-$
in the $\e$-band. Therefore, $\tau >0$ (and $\eta\ge 0$). Due  to
Lorentz-invariant form of diagrams in $k_+,k_-$ we have
$\om_+-\om_- =0$. It follows from the expression (\ref{B11b}) for
a $\Pi$-line propagator that the $\mu$  can  be  counted  as  the
number of $\Pi$-lines in the  subdiagram  taken  with  the  minus
sign. Therefore, one has $ -\mu + \eta >0$ (and,  hence,  $\s>0$)
if  at  least  one  of  $\Pi$-lines  is  present.   Thus,    only
configurations without $\Pi$-lines, i.~e. at  $\m=\et=\s=0$,  can
contribute to the difference between LF and  covariant  diagrams.
The absence of the $\Pi$-lines  means that all external lines of the diagram
are attached to only one vertex. We call such diagrams generalized tadpole.
The logarithmic corrections, that are mentioned before the eq-n
(\ref{B11}) and can arise in general, are present, in fact,
only if they depend on external momenta. This is related with
the fact that
owing to the Lorentz invariance the argument of the logarithm can 
contain the $\e$ only if it contains also a $p_-$-component of external
momenta. However the generalized tadpole diagrams do not depend on external
momenta, therefore logarithmic corrections are absent.

We see that the difference between the Lorentz-covariant and LF 
Feynman integrals in considered theory with polynomial interaction is 
caused only by generalized tadpoles.

\end{document}